\documentclass[usenatbib]{mn2e}
\usepackage{times,graphics,amssymb,epsfig,color}

\newcommand{\ber}{\begin{eqnarray}}
\newcommand{\eer}{\end{eqnarray}}

\def\labequn #1{\label{eq:#1}}
\def\labfig #1{\label{fig:#1}}
\def\labsecn #1{\label{sec:#1}}
\def\labsubsecn #1{\label{subsecn:#1}}
\def\labsubsubsecn #1{\label{subsubsecn:#1}}

\def\equn #1{Equation~\ref{eq:#1}}
\def\eqn #1{Eqn~\ref{eq:#1}}
\def\dequn#1#2{Equations~{\ref{eq:#1}}~and~{\ref{eq:#2}}}
\def\tequn#1#2#3{Equations~{\ref{eq:#1}},~{\ref{eq:#2}}~or~{\ref{eq:#3}}}
\def\fig #1{Figure~\ref{fig:#1}}
\def\dfig#1#2{Figures~\ref{fig:#1}~and~{\ref{fig:#2}}}
\def\tfig#1#2#3{Figures~\ref{fig:#1},\ref{fig:#2}~and~{\ref{fig:#3}}}
\def\secn #1{Section~\ref{sec:#1}}
\def\dsecn#1#2{Sections~{\ref{sec:#1}}~and~{\ref{sec:#2}}}
\def\subsecn #1{Section~\ref{subsecn:#1}}
\def\subsubsecn #1{Section~\ref{subsubsecn:#1}}

\def\etal{et al.\ }
\def\eg{e.\,g.\,}
\def\ie{i.\,e.\,, } 

\def\unit #1{\,{\rm #1}}
\def\nh{N_{\rm H}}
\def\cmsqi{\unit{cm^{-2}}}
\def\kel{\unit{K}}
\def\onlyten#1{10^{#1}}

\def\ev{\unit{eV}}
\def\kev{\unit{keV}}

\def\zsol{Z_{\odot}}

\title[Soft spectral components \& warm absorber]
{The influence of soft spectral components on the structure and stability of warm 
absorbers in AGN}
\author[Chakravorty \etal]
{
Susmita Chakravorty$^{1,2,3}$
\thanks{E-mail: schakravorty@head.cfa.harvard.edu (SC); rmisra@iucaa.ernet.in (RM); 
akk@iucaa.ernet.in (AK); elvis@head.cfa.harvard.edu (MA);
gary@pa.uky.edu (GF)},
Ranjeev Misra$^{1*}$, Martin Elvis$^{3*}$, Ajit K. Kembhavi$^{1*}$, 
Gary Ferland$^{4*}$ \\
$^{1}$IUCAA, Post Bag 4, Ganeshkhind, Pune 411 007, India; \\
$^{2}$Astronomy Department, Harvard University, Cambridge, MA 02138, USA; \\
$^{3}$Harvard-Smithsonian Center for Astrophysics, Cambridge, MA 02138, USA; \\
$^{4}$Department of Physics and Astronomy, University of Kentucky, Lexington, KY 40506, USA.
}

\begin{document}
\maketitle


\begin{abstract}

The radiation from the central regions of active galactic nuclei, including
that from the accretion disk surrounding the black hole, is likely to peak in
the extreme ultraviolet $\sim 13 -100 \ev$. However, due to Galactic
absorption, we are limited to constrain the physical properties, i.e. the black
hole mass and the accretion rate, from what observations we have below $\sim 10
\ev$ or above $\sim 100 \ev$. In this paper we predict the thermal and
ionization states of warm absorbers as a function of the shape of the
unobservable continuum. In particular we model an accretion disk at $kT_{in}
\sim 10 \ev$ and a {\it soft excess} at $kT_{se} \sim 150 \ev$. The warm
absorber, which is the highly ionized gas along the line of sight to the
continuum, shows signatures in the $\sim 0.3 - 2 \kev$ energy range consisting
of numerous absorption lines and edges of various ions, some of the prominent
ones being H- and He-like oxygen, neon, magnesium and silicon. We find that the
properties of the warm absorber are significantly influenced by the changes in
the temperature of the accretion disk, as well as by the strength of the {\it
soft excess}, as they affect the optical depth particularly for iron and
oxygen.  These trends may help develop a method of characterising the shape of
the unobservable continuum and the occurrence of warm absorbers.

\end{abstract}




\section{Introduction}
\labsecn{introduction}

Accretion of matter onto a supermassive black hole and the conversion of the
gravitational energy into radiation, via an accretion disk, is the main source
of energy output in active galactic nuclei (AGN). The radiation from AGN is
likely to peak in the extreme ultraviolet (EUV) $\sim 10 - 100 \ev$
\citep{lynden-bell69, pringle73, shields78, shakura73, netzer85}.  However,
Galactic absorption of EUV light results in a `blind spot' in the energy range
$13 - 100 \ev$ ($\sim 912 - 50 \AA$, marked as ``unobserved'' in
\fig{SedCompare}) introducing an uncertainty in the shape of the spectral
energy distribution (SED) of AGN, just where it is expected to peak. Attempts
have been made to reconstruct the SED from observations available below $\sim
10 \ev$ and above $\sim 100 \ev$ \citep[][and references
therein]{siemiginowska95, sobolewska04a, sobolewska04b}.  This unobserved part
of the SED is most effective at ionizing gas (which is the very reason it also
gets absorbed by the Galaxy). Hence the continuum will influence the
intervening line of sight gas, local to the AGN, and the resultant absorption
and emission line strengths of the various ions seen in far ultraviolet (FUV)
and the Balmer continuum can be used to constrain the SED, as discussed by
\citet{netzer87}.   

At the high energy end of the `blind spot', i.e. in the soft X-ray (0.1 - 10
$\kev$), observations of most AGN reveal that, for a majority, the SED from 2 -
10 keV is well approximated by a power-law $f(\nu) \sim \nu^{-\alpha}$ with
spectral indices $\sim 0.8$. The join between the EUV ($2500 \AA$) and the soft
X-ray ($2 \kev$), $\alpha_{OX}$, is steeper, with typical values of $\sim 1.2$
for Seyfert 1 galaxies \citep{netzer93}.  Hence extending the X-ray power-law
to the ultraviolet (UV) cannot explain the observed flux (but see
\citet{laor97}). An ionizing continuum with two power-law components, a steep
one for the join between EUV and soft X-ray, and a flatter one for the soft
X-ray SED, can represent the overall shape of the AGN spectra in the energy
range $3 \ev - 2 \kev$. 

However, such an SED is not a physical model and does not resemble certain
specific features expected in this energy range. In particular, the proposed
accretion disk, whose innermost stable orbit should have a temperature of $\sim
13 \ev$, for a black hole of mass $10^8 M_{\odot}$ and an accretion rate of
$\dot{m} / \dot{m}_{Edd} = 0.1$ \citep{shakura73,frank02}, is not included.
Another spectral component which would remain unaccounted for by a simple two
power-law continuum is the {\it soft excess} which is seen in most AGN below $2
\kev$. The {\it soft excess} could be a Comptonised disk component
\citep{czerny87}, an additional quasi-blackbody \citep{crummy06,
korista97,ross93,ross05} or due to relativistic broadened absorption features
\citep{gierlinski04}. Such spectral components are likely to have important
effects on the nature of the absorbing gas along the line of sight towards the
central engine of the AGN.

In the 0.3 - 1.5 $\kev$ soft X-ray spectra of about half of all Seyfert 1
galaxies \citep{nandra94,reynolds97,george98} and quasars \citep{piconcelli05}
one can find signatures of photoionized gas called ``warm absorbers''
(hereafter WAs, \citet{halpern84}). These signatures are, often, absorption
lines and edges from highly ionized species, such as OVII, OVIII, FeXVII, NeX,
CV and CVI \citep{collinge01, kaastra02, kinkhabwala02, blustin03, krongold03,
netzer03, turner04}.  The typical column density observed for the gas is $\nh
\sim \onlyten{22\pm1}\cmsqi$. The ionization parameter $\xi$ of the WA spans a
range $\sim 10 - 1000 \,\, \rm{erg\,cm\,s^{-1}}$ corresponding to gas
temperatures of $10^4 - 10^{6.5}$ K. Some authors claim that the WA is made up
of discrete thermodynamic phases in near pressure equilibrium \citep[][and
references therein]{andrade-velazquez10}, while others \citep{ogle04,
steenbrugge05, behar09} believe that the gas has a continuous distribution of
temperature and pressure. 

In this temperature range heating and cooling processes are respectively
dominated by photoionization and recombination of higher ionization states of
heavier elements. These processes result in local regions of thermal stability
in this otherwise unstable temperature range \citep{krolik81, gehrels93,
hess97}. \citet{chakravorty09} showed that the nature of the WA is strongly
influenced by the chemical composition of the absorbing gas. It was shown that
the abundance of iron and oxygen, which have important atomic transitions in
the sub-$\kev$ energy range are particularly important. 

To obtain the equilibrium conditions for a gas with an assumed set physical
conditions, we have used the publicly available photoionization code {\small
CLOUDY}\footnotemark version C07.02 (hereafter C07), see \citet{ferland98}.
\footnotetext{URL: http://www.nublado.org/ } {\small CLOUDY} calculates the
ionization equilibrium conditions by solving the energy and charge conservation
equations under the assumption that all the atomic processes have had time to
reach a steady state.  Any stable photoionised gas will lie on the thermal
equilibrium curve (hereafter referred to as the {\it stability curve}) where
heating balances cooling. The stability curve of temperature ($T$) against
pressure ($\xi/T$), where $\xi$ is the ionization parameter (see below for
definition), is often used to study the nature of the WA.  Gas lying off the
stability curve will heat or cool until reaching the curve. If the curve has
kinks that produce multiple stable values at fixed $\xi/T$, then the WA can
have multiple temperature phases in pressure equilibrium. The shape of the
stability curve depends on the SED of the ionizing continuum and the chemical
abundance of the gas \citep{reynolds95, krolik01, komossa00, komossa01,
chakravorty08, chakravorty09}. 

In this paper we investigate the behaviour of the stability curve as a function
of the {\it disk blackbody} and {\it soft excess} both to understand the role
of these spectral components on the nature of the WA, and to examine whether
the observable parameters of the WA can diagnose the shape of the AGN continuum
in the EUV `blind spot'. In \secn{continuum} we describe the different models
of the AGN continuum that we are interested in and use these SEDs to study
their effect on the stability curves in \secn{Scurves}. We investigate the
causes of the variations in the stability curves in \secn{HeatingCooling} by
studying the heating and ionization fractions of the elements and ions which
are responsible for determining the nature of the WA. \secn{OtherPars} examines
the relevance of a few other physical parameters, like the X-ray slope of the
ionizing continuum and the abundance of the absorbing gas, \citep[also
discussed in][]{chakravorty09} in the context of the present study. The
multi-phase nature of the WA is discussed in \secn{Multi-phase} and we conclude
our results in \secn{Conclusion}. 


\section{The AGN continuum}
\labsecn{continuum}

The 2 - 10 keV X-ray spectra of most of the observed active galactic nuclei
(AGN) can be modeled satisfactorily using a power-law $f(\nu) \sim
\nu^{-\alpha}$ where the spectral index $\alpha$ for most AGN lie in the range
$0.7<\alpha<0.9$ \citep{wilkes87, grupe06, lopez06}. The optical depth $\tau =
1$ for photons at $0.2 \kev$ with a typical Galactic hydrogen column density of
$N_{\rm{H}} = 3 \times 10^{20} \cmsqi$. Extinction is a steeply rising function
of column density, \eg $\tau = 2$ at $0.2 \kev$ for $N_{\rm{H}} = 5 \times
10^{20} \cmsqi$. Thus, we lack observational constraints on the evolution of
the X-ray power-law at lower energies. However, for most AGN, the UV flux is
found to be much higher than what a simple extrapolation of the X-ray power-law
to lower energies would predict \citep{elvis94, zheng97, shang05}, but see
\citet{laor97} for exceptions. The slope $\alpha_{OX}$, defined as
\begin{equation}
\alpha_{OX} = -0.384 \log\left[\frac{f(2 keV)}{f(4.7 \ev  = 2500 \AA)}\right]
\labequn{alphaOX}
\end{equation}
by \citet{tananbaum79}, is conventionally used to parametrize the nominal
power-law between the UV and soft X-ray bands. The observed range is $1
\lesssim \alpha_{OX} \lesssim 2$ \citep[][and references therein]{stalin09,
green09}. 

The overall shape of the AGN continuum in the energy range of $\sim 4.7 \ev
(2500 \AA) - 10 \kev$ can, thus, be represented as a sum of two power-law
components
\begin{equation}
f(\nu) \sim \left[ \nu^{-\alpha} + A_1 \nu^{-\alpha_s} \right] e^{-\frac{\nu}{\nu_{max}}}, ~~~ \rm{for} \,\, E(= h\nu) \ge 4.7 \ev,
\labequn{2powerlaw}
\end{equation}
where $\alpha_s (> 0)$ is the spectral index of a steep soft component and
$A_1$ is the relative normalization factor to realize the desired values of
$\alpha_{OX}$ (dotted-and-dashed curve in \fig{SedCompare}.).
\citet{chakravorty09} had used such an SED with $E_{max} = h\nu_{max} = 200
\kev$, and this value is maintained throughout this paper, as well. Results
from the Swift/BAT (Burst Alert Telescope) hard X-ray sky survey
\citep{tueller08} show that, of the brightest few dozen AGN, for which
$E_{max}$ can be determined, the cut-off spans the range 50 to 450 $\kev$. In
\citet{chakravorty09} we have shown that in the range $50 \kev < E_{max} < 400
\kev$, the variation in $E_{max}$ does not have any significant effect on the
nature of the warm absorber. For energies lower than $4.7 \ev$
\citet{chakravorty09} followed the cut-off scheme 
\begin{eqnarray}
f(\nu) & \sim & \nu^{-0.5} ~~~ \rm{for} \,\ 2.8 \,\, \le E(= h\nu) < 4.7 \ev \nonumber \\
& \sim & \nu^{-1.0} ~~~ \rm{for} \,\ 0.12 \le E(= h\nu) < 2.8 \ev \nonumber \\
& \sim & \nu^{2.5} ~~~ \rm{for} \,\ E(= h\nu) < 0.12 \ev 
\labequn{Emin}
\end{eqnarray}
This scheme is similar to that described by \citet{mathews87}. However, a SED
described by \dequn{2powerlaw}{Emin} may sometimes be an inadequate description
for the AGN ionizing continuum because it fails to represent the signature of
the accretion disk at $\sim 10 \ev$ or the {\it soft excess} component.

\begin{figure}
\begin{center}
\includegraphics[scale = 1, width = 8 cm, trim = 0 50 30 0, clip, angle = 0]{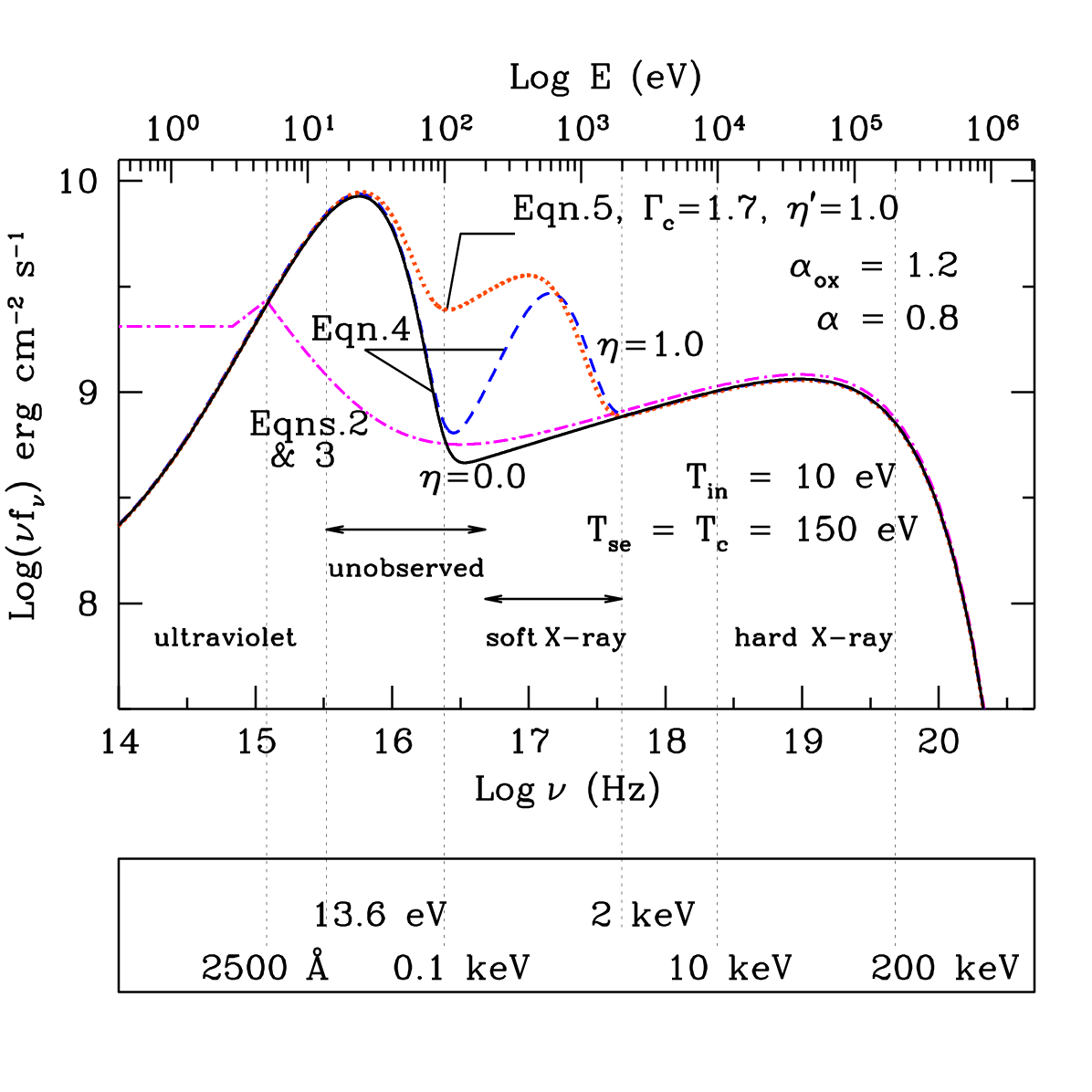}
\caption{Comparison between various ionizing continua : the broken power-law 
(\equn{2powerlaw}, dotted-and-dashed line) and more realistic SEDs having {\it disk blackbody} and 
{\it soft excess} components (\dequn{GenSpectra}{CompSe}). For all the curves $\alpha = 0.8$ 
and $\alpha_{OX} = 1.2$. Using \equn{GenSpectra} we draw the solid curve with $\eta = 0.0$ 
and the dashed curve with $\eta = 1.0$, while the dotted curve is drawn using 
\equn{CompSe} with $\eta^{\prime} = 1.0$. The temperature of the innermost ring of the 
accretion disk is $T_{in} = 10 \ev$ and temperature corresponding to the {\it soft excess} 
is $T_{se} = T_c = 150 \ev$. See text for further details.}  
\labfig{SedCompare}
\end{center}
\end{figure}

\subsection{Disk blackbody}
\labsubsecn{SED_Dbb}

Multiwavelength observations suggest that the AGN continua peak in the EUV
energy band and emission here usually dominates the quasar luminosity
\citep{neugebauer79, shields78, malkan82, elvis86, czerny87, mathews87, laor90,
siemiginowska95, zheng97, sobolewska04a, sobolewska04b}.  This spectral
component, often referred to as the ``Big Blue Bump'', is considered to be the
signature of the presence of an accretion disk as discussed by
\citet{lynden-bell69, pringle73, shields78, shakura73}.

According to the standard theory of accretion disks by \citet{shakura73}, the
emission from the disk can be modeled as a sum of local blackbody radiations
emitted from annuli of the disk at different radii. The temperature of all the
subsequent rings can be estimated from the theory if the temperature $T_{in}$
of the innermost ring of the accretion disk is known, which in turn is related
to the one-fourth power of the mass of the central black hole and to the fourth
power of its accretion rate. Thus for the same accretion rate, $T_{in}$ changes
by only a factor of 3 while the black hole mass spans a range of about two
order of magnitude. As such, the resultant shape of the ``Big Blue Bump'' peaks
between $\sim 10 - 100 \ev$ for the range of black hole masses and accretion
rates typical of Seyfert galaxies.

Studies of high redshift quasars show that their UV-EUV SEDs are consistent
with the standard theory of accretion disks \citep{bechtold94, page04a,
page04b, shang05}. However, typically, the quasars have higher mass
\citep[$10^8 \lesssim \rm{M_{BH}} \lesssim 5 \times 10^{10}
\rm{M_{\odot}}$][]{shang05,bechtold94} and higher accretion rates
\citep{bechtold94}. Thus, for the same accretion rates, we can expect the ``Big
Blue Bump'' in quasars to be peaking at lower energies than that in Seyferts
which would typically result in steeper $\alpha_{OX} \sim 1.8$
\citep{bechtold94} for the quasars, as compared to 1.2 \citep{netzer93} for the
Seyferts.

A standard model for the spectral component from the accretion disk is
available as {\it disk blackbody} \citep[][]{mitsuda84, makishima86} in
XSPEC\footnote{http://heasarc.gsfc.nasa.gov/docs/xanadu/xspec/}
\citep{arnaud96}. We have used version 11.3 of XSPEC to generate the {\it disk
blackbody} spectral component $f_{dbb}(\nu)$ to be used in the construction of
the realistic AGN continuum (the third term in \dequn{GenSpectra}{CompSe}). 
The solid black line in \fig{SedCompare} shows a SED which includes a ``Big
Blue Bump'' with $T_{in} = 10 \ev$. In this paper we explore ``Big Blue Bumps''
in the range $10 \le T_{in} \le 30 \ev$ corresponding to $10^{8.47} \le M_{BH}
\le 10^{6.56} M_{\odot}$ for an accretion rate of $\dot{m} / \dot{m}_{Edd} =
0.1$. This range of accretion disks parameters pertain to Seyfert galaxies or
low mass and low accretion rate quasars. We do not explore the parameter range
for higher mass quasars because for them $T_{in}$ would move to lower energies
and such {\it disk blackbodies} are less likely to influence the nature of the
WA.


\subsection{Soft excess}
\labsubsecn{SED_Se}

X-ray observations of AGN with ROSAT and XMM-Newton often show that if the 1 -
10 keV power-law is extended to lower energies to fit the observed spectra of
type 1 AGN, some unaccounted for excess intensity is usually seen at $E < 1
\kev$ \citep{elvis85, brinkmann92, buehler95, pounds02}.  This excess has come
to be known as the {\it soft excess} component. A blackbody with temperature
$T_{se} \sim 100 - 200 \ev$ (i.e. peaking at $\sim 282 - 564 \ev$) is often a
good fit to the {\it soft excess} \citep[][and references therein]{matsumoto04,
piconcelli05, porquet04, vignali04}.  The same authors show that the ratio of
the {\it soft excess} luminosity to power-law luminosity, usually between 0.1 -
10 keV varies from object to object from 0.04 in Mkn 304 \citep{piconcelli05} to
$\gtrsim 1.0$ in Ark 564 \citep{vignali04}.  The simple `sum of blackbodies'
model for the accretions disk is a satisfactory qualitative representation of
the UV SED with a range of parameter values to cover the observed AGN
properties. However, all AGN disks, having supermassive black-holes at their
centres, are too cold to reach soft X-rays at $\sim 0.5 \kev$. More
sophisticated modifications of this model \citep{czerny87, korista97} or an
additional spectral component is required to explain the {\it soft excess}.


\subsubsection{Blackbody soft excess}
\labsubsubsecn{SED_BbSe}

A theoretical representation of the {\it soft excess} as a blackbody is a
simplified version of the \citet{ross93, ross05} model for the {\it soft
excess} component which owes its origin to the reflection of the power-law
component of the AGN spectrum from the accretion disk. Thus, in this model the
X-ray power-law photons are reprocessed, instead of the ones from the accretion
disc. 

The general SED for the ionizing continuum including a blackbody {\it soft
excess component} can be given as 
\begin{eqnarray}
f(\nu) \,\, \sim \,\, \big[\,\, \{\, \nu^{-\alpha} \, + \, \eta \frac {2 \, \pi \,h} {c^2} \, \frac {\nu^3} {\exp(h\,\nu/k_B\,T_{se})\, - \,1} \, \} \nonumber \\
+ \,\, A_2 f_{dbb}(\nu, T_{in}) \,\, \big] \,\, e^{-\frac{\nu}{\nu_{max}}}.
\labequn{GenSpectra}
\end{eqnarray}
The first term in the above equation represents the X-ray power-law with
spectral index $\alpha$. The second term is the {\it soft excess} component, as
a blackbody distribution of photons where $T_{se} = 150 \ev$ is the temperature
of the blackbody and $\eta$ determines the ratio of luminosity in the {\it soft
excess} component to that in the power-law between 0.1 - 10 keV. The third term
in \equn{GenSpectra} is the {\it disk blackbody} component which is
parametrised by $T_{in}$, the temperature of the innermost ring of the
accretion disk. $A_2$ is the normalisation factor to attain the desired value
of $\alpha_{OX}$. Unless otherwise mentioned, throughout this paper, we have
used $\alpha = 0.8$ and $\alpha_{OX} = 1.2$, $E_{max} = h \nu_{max} = 200
\kev$. For numerical convenience we put a cut-off for $f(\nu)$ at $E = h\nu =
0.12 \ev$ below which the flux drops off as $\nu^{2.5}$. 

In \fig{SedCompare} both, the black solid SED and the blue dashed SED are drawn
using \equn{GenSpectra}. For both SEDs, $T_{in} = 10 \ev$ so that the peak of
the diskblackbody lies at $\sim 30 \ev$. The solid line represents a SED which
does not have any {\it soft excess} component, \ie $\eta = 0$. On the other
hand, the blue dashed SED with $\eta =1.0$ has a moderately strong {\it soft
excess}. See further discussions in \subsubsecn{Scurve_BbSe}


\subsubsection{Comptonisation by $\sim 150 \ev$ plasma}
\labsubsubsecn{SED_CompSe}

\citet{korista97} discuss the puzzle that many quasars which show soft ionizing
continuum, would have an insufficient number of photons at the 54.4 eV He II
edge to generate the observed strength of He II emission seen in the same
objects.  To explain the strength of the He II line in Mrk 335,
\citet{korista97} invoked a ``double peaked'' UV-EUV continuum where the second
``bump'' peaks at $\sim 54 \ev$ and contains energy comparable to the classical
UV bump at lower energies ($\sim 10 \ev$). 

A model, where the photons from the accretion disk are upscattered by inverse
Comptonization due to energetic electrons, does a better job of qualitatively
satisfying the required strength of the UV-EUV SED for objects like Mrk 335.
\citet{czerny87} have shown that the models which account for the electron
scattering of the accretion disk photons are better than the simple `sum of
blackbodies' model. Similar thermal Comptonization models have been worked out
by \citet{lightman87, coppi92, haardt93, coppi99} and \citet{beloborodov99}
among others.

The thermal Comptonization model of \citet{lightman87} has been used by
\citet{zdziarski96} and extended by \citet{zycki99} and their model is included
in XSPEC 12.5 \citep{arnaud96} as {\it nthcomp}. We have used {\it nthcomp},
where we have assumed that the seed photons coming from the accretion disk
(modeled as {\it disk blackbody}) are reprocessed by the thermal plasma to
generate sufficient photons at sub-keV to mimic the {\it soft excess}, usually
observed in AGN. The high energy cut-off for the resulting {\it soft excess}
feature is  parametrised by the electron temperature $T_c$, whereas the low
energy rollover is dependent on the effective temperature of the seed photons
from the accretion disk, which in this case is parametrised by $T_{in}$.
Between the low and high energy rollovers the shape of the spectrum is not
necessarily a power law, but can be parametrised by an asymptotic power law
index $\Gamma_c$ which would physically be determined by the combination of
electron scattering optical depth and electron temperature, i.e. by the Compton
y-parameter \citep{rybicki86}. \citet{beloborodov99} had shown that there is a
simple relation $\Gamma_c \approx \frac{9}{4} y^{-2/9}$ between the power-law
index and the Compton y-parameter. Thus, in our model, $\Gamma_c$, used as an
input, gives a measure of the extent of Compton reprocessing; i.e. larger the
value of $\Gamma_c$, lesser is the number of photons reprocessed from the {\it
disk blackbody} component to the high energy photons at $\sim T_c$. 

The SED with this alternative model for the {\it soft excess} $f_c(\nu)$,
generated by {\it nthcomp}, can be written as
\begin{eqnarray}
f(\nu) \,\, \sim \,\, \big[ \,\, \{\, \nu^{-\alpha} \, + \, \eta^{\prime} f_c(\nu, T_{in}, T_c, \Gamma_c) \,\} \nonumber \\
+ \,\, A_2 f_{dbb}(\nu, T_{in}) \,\, \big] \,\, e^{-\frac{\nu}{\nu_{max}}},
\labequn{CompSe}
\end{eqnarray}
where $\eta^{\prime}$ is the ratio of luminosity of the {\it soft excess}
component for $E (= h\nu) \ge 0.3 \kev$ to the luminosity of the power-law
component in the energy range $0.1 - 10 \kev$. The orange dotted curve in
\fig{SedCompare} is drawn using \equn{CompSe} with $T_{in} = 10 \ev$, $T_c =
150 \ev$ $\eta^{\prime} = 1.0$ and $\Gamma_c = 1.7$. Note that at $E = 54.4 \ev
\,\, (\rm{i.e.} \,\, \log \nu = 16.12)$ the orange dotted SED has 1.4 times
higher flux than the blue dashed SED due to a blackbody {\it soft excess} and
this factor grows to 13.1 at $E = 100 \ev$. Thus SEDs like the orange dotted
line are better suited to explain observations of objects like Mrk 335 or Mrk
478. See further discussions on this issue in \subsubsecn{Scurve_CompSe}.

\citet{chakravorty09} showed that the WA temperature range, $10^{4.5} < T <
10^{6.5} \kel$, is strongly influenced by the chemical composition of the
absorbing gas, particularly by the abundance of iron and oxygen which have
important atomic transitions in the sub-keV energy range where the {\it soft
excess} component is likely to have maximum effect. In the following sections
we shall extensively investigate the effect of a SEDs, given by
\dequn{GenSpectra}{CompSe}, on the nature of the WA, as a function of the shape
of the accretion disk component parametrised by the value of $T_{in}$, and the
strength of the {\it soft excess} feature parametrised by $\eta$ or $\Gamma_c$
and $\eta^{\prime}$.



\section{Stability curves analysis}
\labsecn{Scurves}

Studies of WA variability in response to continuum changes show that it is
reasonable to assume the WA to be in ionization and thermal equilibrium as
observed for NGC 985 \citep{krongold05a}, NGC 3783 \citep{krongold05b}, NGC
5548 \citep{andrade-velazquez10}.  The thermal and ionization equilibrium is
governed by heating due to photoionization and cooling due to line emission and
collisional recombination (radiative and dielectronic).

\begin{figure*}
\begin{center}
\includegraphics[scale = 1, width = 18 cm, trim = 20 390 20 125, clip, angle = 0]{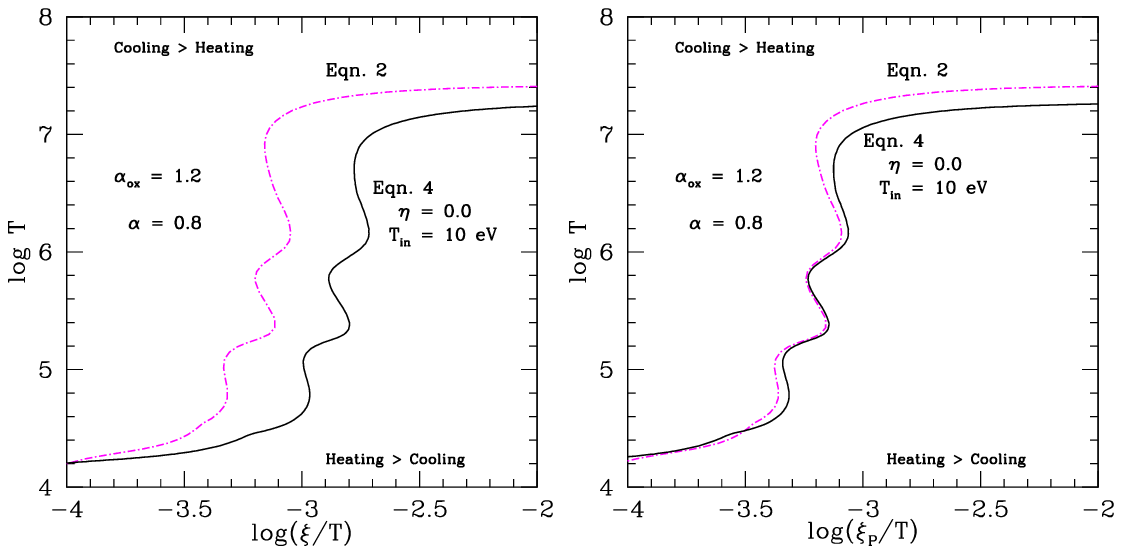}
\caption[Stability curves demonstrating the advantage of using $\xi_P$ as
compared to $\xi$] {Stability curves demonstrating the advantage of using
$\xi_P$ as compared to $\xi$. The dotted-and-dashed stability curves in both
the panels are drawn using the broken power-law continuum (\equn{2powerlaw}),
whereas the solid ones are generated for a SED given by \equn{GenSpectra} with
$\eta = 0.0$ and $T_{in} = 10 \ev$. Curves drawn with $\xi$ (left panel) are
separated in the phase space even if the WA in both the cases have similar
nature. On the other hand, the curves drawn using the alternative definition,
$\xi_P$, of the ionization parameter (right panel), are almost identical
showing the true nature of the WAs. See text in \subsecn{XiP} for further
details.}  
\labfig{XiXiPCompare}
\end{center}
\end{figure*}

We model the WA as an optically thin, plane parallel slab of Solar metallicity
gas {\bf (as given by \citet{allendeprieto01, allendeprieto02} for C and O, by
\citet{holweger01} for N, Ne, Mg, Si and by \citet{grevesse98} for the
remainder of the first thirty elements)} with column density $\nh = 10^{22}
\cmsqi$.  The absorbing gas is assumed to be illuminated by an ionizing
continuum given by \equn{GenSpectra} or \equn{CompSe} and the ionization state
of the gas can be described by specifying the ratio of the ionizing photon flux
to the gas density through an {\it ionization parameter} (see \equn{ionparm}). 

\subsection{Normalising the ionisation parameter}
\labsubsecn{XiP}

The photoionization state of the WA can be parametrised by the ionization
parameter which is the ratio of the ionizing photon flux to the gas density
\citep{tarter69}
\begin{equation}
\xi \,\, = \,\, L_{\rm{ion}}/n_{\rm{H}}\,R^2 \,\, = \int\limits_{h\nu_0 = 13.6 \ev}^{13.6 \kev} \frac{L_{\nu}}{n_H R^2} \, d\nu~~~~~~~\unit{[erg\,cm\,s^{-1}]},
\labequn{ionparm}
\end{equation}
where $L_{\rm{ion}}$ is the luminosity between $13.6 \ev - 13.6 \kev$ (i.e $1 - 10^3$ Rydberg). Hence 
$\xi/T \sim L/pR^2$, $p$ being the gas pressure in the WA. 

This definition has a certain drawback which is demonstrated by the
stability curves on the left panel of \fig{XiXiPCompare}. The dotted-and-dashed
curve is for a gas illuminated with the broken power-law ionizing continuum
(\equn{2powerlaw}) and the solid curve is generated with a more realistic SED
given by \equn{GenSpectra} having an accretion disk component with $T_{in} = 10
\ev$, but no {\it soft excess} ($\eta = 0.0$). From \fig{SedCompare} we can see
that the SED including the accretion disk component (the solid black line) has
$\sim 50$ times more photons at $13.6 \ev$ (1 Rydberg) than the
dotted-and-dashed spectra (\equn{2powerlaw}), although they have similar flux
at 100 eV. This would result in very different values of $\xi$ in the two cases
but very similar WAs, since the WA properties are determined by the photon
distribution in soft X-ray ($E \gtrsim 100 \ev$) and not by photons with energy
$E << 100 \ev$. The {\it kinks} in any stability curve are a result of the
interplay between the various heating and cooling agents responsible for
maintaining the gas at a state of thermal equilibrium \citep{chakravorty08,
chakravorty09}. The stability curves on the left panel of \fig{XiXiPCompare}
have almost identical shape for $4.4 < \log T < 6.5$, indicating that they have
WAs with identical thermal properties and state of ionization. However, they
are separated from each other, by a mere horizontal shift in $\log(\xi/T)$
because of the different $\xi$ values predicted by the two different SEDs.
Thus, $\xi$ is a relatively poor parametrization of the nature of the WA. Such
a problem with the standard definition of the ionization parameter has been
acknowledged by other authors as well. For example, \citet[and references
therein]{chelouche05} use an ionization parameter $U_x$ which considers the
ionizing flux only between $540 \ev \,\,\rm{to} \,\, 10 \kev$.

To circumvent this problem we use a normalisation scheme, appropriate for
this paper, as described in the following. All the SEDs considered in this
paper (except for ones in \subsecn{Alpha}) have a common power-law in addition
to the diskblackbody and/or the {\it soft excess} component. Hence we define
\begin{equation}
\xi_P \,\, = \,\, \frac{L_P}{n_{\rm{H}}\,R^2} \,\, = \frac{1}{n_{\rm{H}}\,R^2} \left[ \,\, 4 \pi R^2 \, \int\limits_{h\nu_0 = 13.6 \ev}^{200 \rm{keV}} \nu^{-\alpha} \, d\nu \,\, \right]
\labequn{ionparmP}
\end{equation}
where $L_P$ is a constant luminosity due to the soft X-ray power-law component
$f(\nu) \sim \nu^{-\alpha}$ (with $\alpha = 0.8$) in the energy range $E = h\nu
= 13.6 \ev - 200 \kev$. Thus, $\xi_P$ is the ratio of the ionizing flux, due
only to the power-law component in the energy range $13.6 \ev \,\,\rm{to} \,\,
200 \kev$, to the gas density. For any SED given by
\tequn{2powerlaw}{GenSpectra}{CompSe} there would be a unique factor
$\xi/\xi_P$ for the same $n_{\rm{H}}\,R^2$. The stability curves corresponding
to the ionizing continua given by \tequn{2powerlaw}{GenSpectra}{CompSe} are
calculated by {\small CLOUDY} taking the entire SEDs into account, following
which their x-axis ($\log(\xi/T)$) is divided by the unique (to each SED)
factor $\xi/\xi_P$ generating a normalised stability curve in the $\log T$ -
$\log(\xi_P/T)$ plane. The advantage of such a normalisation is shown in the
right panel of \fig{XiXiPCompare}, where the stability curves overlap closely,
demonstrating the true physical scenario of two very similar WAs. Note that
these are the same stability curves which are seperated by the horizontal shift
in the left panel, as discussed above. This normalisation scheme is used for
all subsequent figures in the paper except for \fig{Alpha}. In addition to the
SEDs which determine the physics of the respective stability curves, the top
panels of \tfig{DbbSpectraScurves}{BbSeSpectraScurves}{CompSeSpectraScurves}
also include the constant power-law component (magenta, long-and-short-dashed
curve) which is used to normalise the stability curves so that they can be
plotted in terms of $\log(\xi_P/T)$.

It is to be noted that while calculating the thermal and ionization properties
of the WA, {\small CLOUDY} uses the entire SEDs, including all three spectral
components (namely the {\it disk blackbody}, the {\it soft excess} and the
X-ray power-law), described by \dequn{GenSpectra}{CompSe}. The introduction
of $\xi_P$ and normalisation of the stability curves is only for the
convenience of demonstration of the stability curves and their associated
properties. Use of $\xi_P$ merely introduces an overall horizontal shift in
these distributions. Further, note that any given range in $\log(\xi/T)$ would
correspond to an exactly equal range in $\log(\xi_P/T)$ and vice versa.


\subsection{Disk-blackbody}
\labsubsecn{Scurve_Dbb}

The top panel of \fig{DbbSpectraScurves} shows the ionizing continua which
include the {\it disk blackbody} and the power-law components ($\eta = 0$,
\equn{GenSpectra}). As $T_{in}$ is increased from $10 \ev$ to $30 \ev$, the
peak of the flux distribution moves from $\sim 30 - 90 \ev$. Photons at $E
\lesssim 90 \ev$ affect only the lower ionization species of the absorbing gas.
The resulting WA should show differences only in the nature of the lower
ionization phases, and should have very similar higher ionization states. We
shall discuss this issue further in \secn{HeatingCooling}.

\begin{figure}
\begin{center}
\includegraphics[scale = 1, width = 0.55\textwidth, trim = 120 10 50 75, clip, angle = 0]{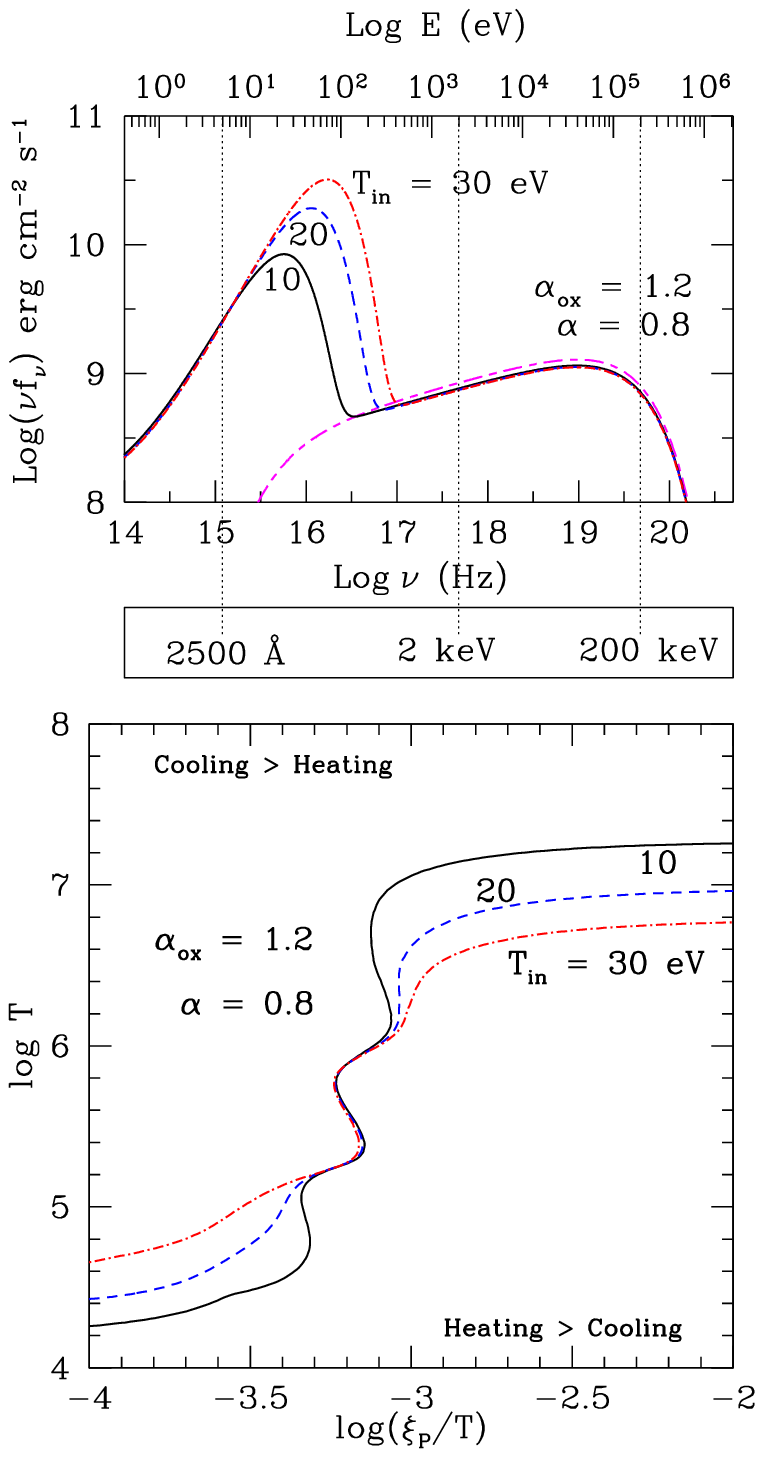}
\caption{The SEDs in the top panel are constituted by two spectral 
components, a X-ray power-law and a EUV {\it disk blackbody}. The curves are for 
different values of the temperature $T_{in}$ of the innermost ring of the accretion 
disk. $\alpha = 0.8$ and $\alpha_{OX} = 1.2$ for all 
the SEDs. We have included the constant power-law (magenta,
long-and-short-dashed curve) which is used to normalise the stability curves so
that they can be plotted in terms of $\log(\xi_P/T)$. Important energy values
including the upper energy cut-off ($E_{max} = 200 \kev$) and the range of
definition for $\alpha_{OX}$ ($2500 \AA$ and 2 \kev) have been marked and
labeled.  The bottom panel shows the stability curves corresponding to the
ionizing continua shown in the top panels.}  
\labfig{DbbSpectraScurves}
\end{center}
\end{figure}

Corresponding stability curves are shown in the bottom panel of
\fig{DbbSpectraScurves}. A hotter {\it disk blackbody} lowers the Compton
temperature branch and increases the temperature of the bottom branch. However,
in the $5 < \log T < {6.5} \kel$ temperature range of the WAs, the stability
curves are independent of the value of $T_{in}$, \ie unaffected by the spectral
component due to the accretion disk. However, the lower temperature ($\log T <
5$) part of the stability curves undergo remarkable changes with the increase
in $T_{in}$; for the same value of $\xi_P$, the gas attains a higher
temperature, the stability curve becomes more stable and any possibility of
multi-phase existence with higher ionization states is lost. See
\secn{Multi-phase} for further discussion on this issue.


\subsection{Soft excess}
\labsecn{Scurve_Se}

\subsubsection{Blackbody {\it soft excess}}
\labsubsubsecn{Scurve_BbSe}

The top panel of \fig{BbSeSpectraScurves} shows the ionizing continua given by
\equn{GenSpectra} with  values of $\eta = 0.0, 1.0, 3.0$, representing
increasing strengths of the {\it soft excess} component for a fixed value of
the blackbody temperature $T_{se} = 150 \ev$. The corresponding stability
curves in the bottom panel show a remarkable enhancement of the stable
$\log(\xi_P/T)$ range from $\sim 0.20 \,\, \rm{through} \,\, 0.41 \,\, \rm{to}
\,\, 0.63$ dex for the $10^5 \kel$ absorber, as the strength of the {\it soft
excess} is increased from $\eta = 0.0 \,\, \rm{through} \,\, 1.0 \,\, \rm{to}
\,\, 3.0$. The $10^6 \kel$ phase remains unchanged through the variation of
$\eta$. In the $\eta = 3.0$ case however, the transition from the $\sim 10^5
\,\, \rm{to} \,\, 10^6 \kel$ phase is smooth i.e. with no distinct unstable
region separating the two temperature regimes. Such a result implies that
stronger the {\it soft excess} component in the spectra, greater is the
probability of finding a $\sim 10^5 \kel$ absorber, since the stable region
then becomes less susceptible to luminosity variations in the AGN. However, the
possibility of the $10^5 \kel$ gas being in pressure equilibrium with the other
WA phases is significantly reduced. We shall return to these points in
\secn{Multi-phase}.

To find a typical value of $\eta$ implied by X-ray observations of AGN we
refer to two examples, namely a typical Seyfert 1 galaxy NGC 5548 and a
narrow line Seyfert 1 galaxy IRAS13349+2438. \citet{andrade-velazquez10}
analysed the 800 ks Chandra grating spectra of NGC 5548 and their best fit
parameters for the soft X-ray continuum is a power-law with $\alpha = 0.6$ and
a blackbody with $T_{se} = 110 \ev$. The relative normalisation of the two
components results in $\eta \sim 1.2$. Similarly, \citet{holczer07} have
analysed the 300 ks {\it Chandra} data for IRAS13349+2438 and have reported the
spectral parameters for the best-fit ionizing continuum comprising of a X-ray
power-law with $\alpha = 0.9$ and a blackbody with $T_{se} = 105 \ev$. From
their results we calculate $\eta$ to be $\sim 2.58$. In both cases the values
of $\eta$ is well within the range considered here (0 - 3).

\begin{figure}
\begin{center}
\includegraphics[scale = 1, width = 0.55\textwidth, trim = 120 0 50 75, clip, angle = 0]{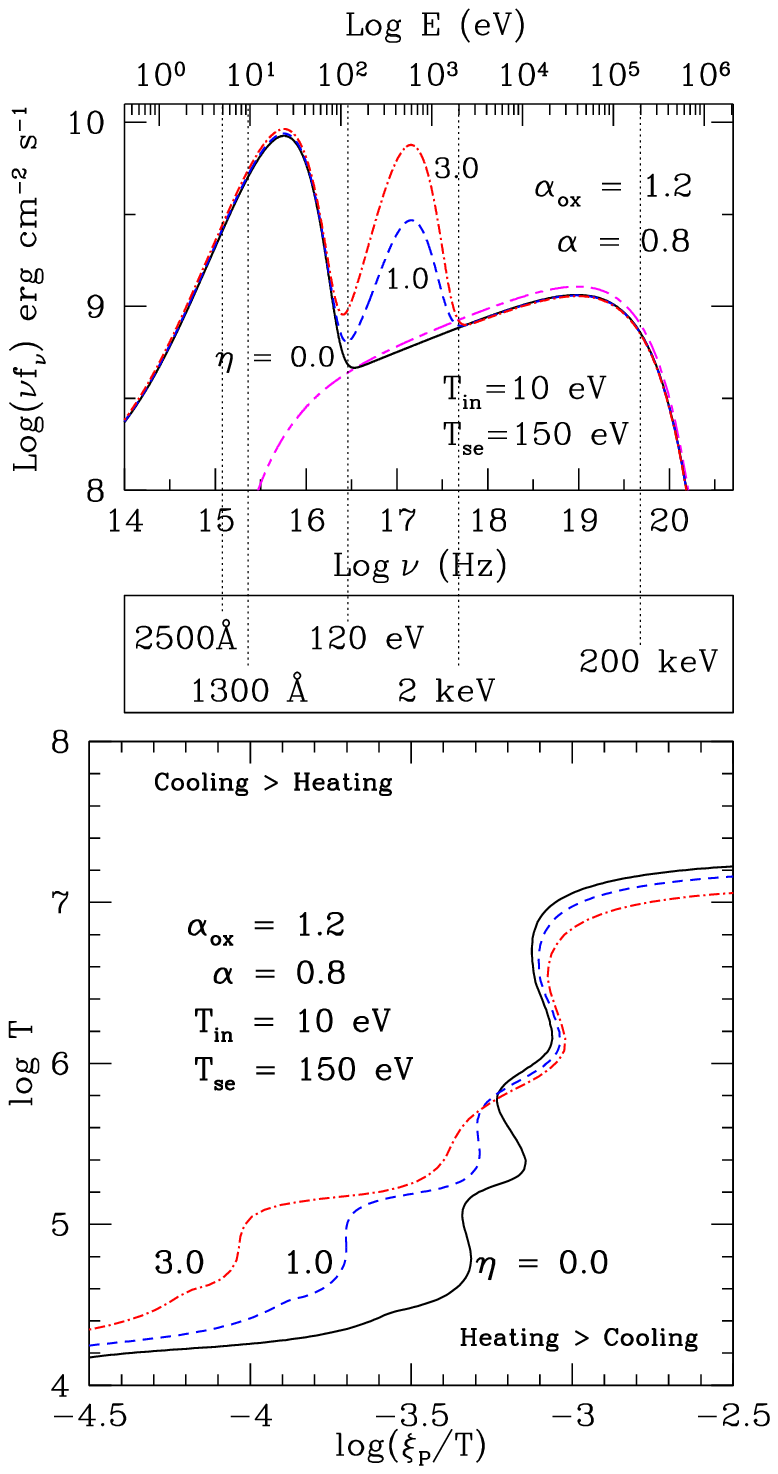}
\caption {\emph{Top panel} : SEDs which have a {\it soft excess} component described 
by a blackbody distribution (see \equn{GenSpectra}) with temperature $T_{se} = 150 \ev$. 
The different SEDs are for different values of $\eta$, \ie for different strengths of the 
{\it soft excess} component. $\alpha = 0.8$ and $\alpha_{OX} = 1.2$ for all the
SEDs. The magenta, long-and-short-dashed curve is the same as in the top
panel of \fig{DbbSpectraScurves}. We have included the energy values $1300
\AA$ and $120 \ev$ to guide the eye in the context of observations of Mrk 478
(see text in \subsubsecn{Scurve_CompSe}) \emph{Bottom panel} : The stability
curves corresponding to the ionizing continua shown in the top panels. With the
growing strength of the {\it soft excess} component, there is significant
increase in the range of $\xi_P$ over which thermally stable warm gas at $\sim
10^5 \kel$ can exist.} 
\labfig{BbSeSpectraScurves}
\end{center}
\end{figure}

The $T_{se}$, observed for a large number of type I AGN, lies in the small
range $100 - 200 \ev$. We have checked whether the influence of the {\it soft
excess} component is modified when the blackbody temperature is varied across
this range. Results showed that the stability curves remain unaffected by such
variations, e.g. the range in $\log(\xi_P/T)$ for the stable $10^5 \kel$ gas
changes by merely 0.06 dex as $T_{se}$ varies from 100 eV to 200 eV.


\subsubsection{Comptonised {\it soft excess}}
\labsubsubsecn{Scurve_CompSe}

\begin{figure}
\begin{center}
\includegraphics[scale = 1, width = 0.55\textwidth, trim = 120 0 50 75, clip, angle = 0]{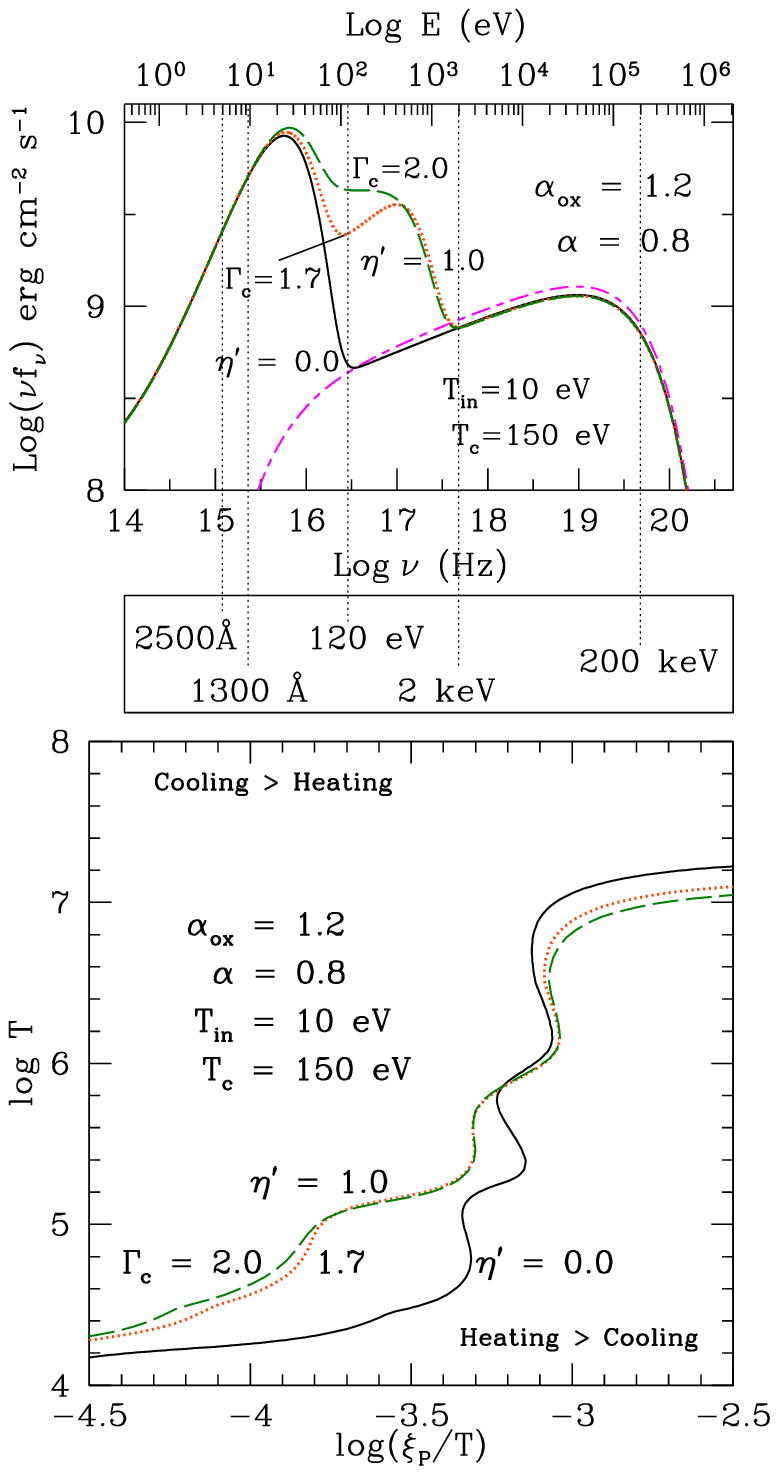}
\caption {\emph{Top panel} : Ionizing continua with {\it soft excess} component due to 
thermal Comptonisation given by \equn{CompSe} with fixed values of 
$T_c = 150 \ev$, $T_{in} = 10 \ev$ and $\eta^{\prime} = 1.0$. The different curves are for 
different values of $\Gamma_c$, namely 1.7 and 2.0. For comparison, we have also drawn the 
continuum with no {\it soft excess} (solid curve). With $\alpha = 0.8$ and 
$\alpha_{OX} = 1.2$, the accretion disk component and the X-ray power-law component are 
same for all the SEDs. The magenta, long-and-short-dashed curve is the
same as in the top panels of \dfig{DbbSpectraScurves}{BbSeSpectraScurves}. We
have included the energy values $1300 \AA$ and $120 \ev$ to guide the eye in
the context of observations of Mrk 478 (see text in \subsubsecn{Scurve_CompSe})
\emph{Bottom panel} : The stability curves corresponding to the ionizing
continua shown in the top panels.}
\labfig{CompSeSpectraScurves}
\end{center}
\end{figure}

The blackbody {\it soft excess} sometimes turns out to be inadequate in
explaining the observations. For example, the SED for Mrk 478 may be 1-3 times
brighter in $\nu L_{\nu}$ at $E \sim 120 \ev$ than it is at $1300 \AA$ ($\sim 10
\ev$) \citep{gondhalekar94, marshall96}. We find that a blackbody {\it soft
excess} cannot satisfy such a condition even with a relatively high
normalisation of $\eta = 3.0$ (top panel, \fig{BbSeSpectraScurves}). We want
to investigate if the alternative description of the {\it soft excess}
component, namely the {\it soft excess} due to thermal Comptonisation, can
account for a SED similar to that seen in Mrk 478, and to see what are its
effect on the WA. 

Spectra generated by \equn{CompSe} with fixed values of $T_c = 150 \ev$,
$T_{in} = 10 \ev$ and $\eta^{\prime} = 1.0$ are drawn in the top panel of
\fig{CompSeSpectraScurves} for different values of $\Gamma_c$.  For comparison,
we have also drawn the continuum with no {\it soft excess}
(\fig{CompSeSpectraScurves} solid curves).  The continuum with $\Gamma_c = 2.0$
satisfies the conditions required by observations of Mrk 478 as mentioned by
\citet{korista97}. 

The influence of a thermal Comptonisation {\it soft excess} component on the
stability curves is shown in the bottom panel of \fig{CompSeSpectraScurves}. We
see again, that the range of $\log(\xi_P/T)$ for the stable $10^5 \kel$ branch
is increased in width from  ${\rm \sim 0.2 \,\, dex \,\, to \,\, 0.48 \,\,
dex}$ as $\eta^{\prime}$ goes from 0 to 1.0, independent of the value of
$\Gamma$, facilitating the presence of a stable WA at these temperatures and
that the different phases of the WA lose the possibility of existing in
pressure equilibrium as the strength of the soft excess component is increased.

The thermal Comptonised and the blackbody models of the {\it soft excess}
result in similar WAs. For example, $\eta = 1.0$ curve (dashed; lower panel of
\fig{BbSeSpectraScurves}) have stable ranges of $\log(\xi_P/T)$ at $10^5 \,\,
\rm{and} \,\, 10^6 \kel$ similar to that of the $\eta^{\prime} = 1.0$ curves
(the dotted and the long-dashed for $\Gamma = 1.7 \,\, \rm{and} \,\, 2.0$
respectively; lower panel of \fig{CompSeSpectraScurves}). The stability curves
with $\eta^{\prime} = 1.0$ do have a slightly smoother rise from the $\sim
10^{4.2} \,\, \rm{to} \,\, 10^5 \kel$ with no distinct intermediate unstable
phase, because they have more flux in the $E = 10 - 100 \ev$ range for the same
{\it soft excess} strength.



\section{Cause of extended stable region : heating agents and ion fractions}
\labsecn{HeatingCooling}

We have seen that the disk blackbody and the {\it soft excess} radiation at
$\lesssim 0.5 \kev$ significantly influence the stability curve, implying that
these components are important for shaping the ionic and thermal state of the
WA (\secn{Scurves}). Changes in the {\it disk blackbody} affect the low
temperature {\it arm} ($\log T \sim 4.5$) of the stability curve, whereas the
{\it soft excess} influences the absorbing gas at $\log T \sim 5.0 \kel$. In
this section we examine the important heating agents and the distribution of
ion fractions of the signature ions in the WA along the stability curve. All
the ionizing continua considered in this section are given by
\equn{GenSpectra}. The effect of the thermal Comptonised {\it soft excess}
gives the same qualitative results as the blackbody {\it soft excess} modeled
by a blackbody (\equn{GenSpectra}, see \labsecn{Scurve_Se}). 


\subsection{Heating}
\labsubsecn{HeatingCooling_Heat}

{\small CLOUDY} works by dividing a gas into a set of thin concentric shells,
referred to as `zones' which have thicknesses that are small enough for the
physical conditions across them to be nearly constant, maintained by
continuously adjusting the physical thicknesses of these shells. For this
sub-section, only, we have constrained the code to perform a single zone
calculation for numerical convenience as we are not concerned with quantitative
rigor, but want to understand the qualitative trends.

\fig{Heating} shows the influence of these two components on the heating
fraction $\Delta H$ of H$^{+0}$, He$^{+1}$, Fe and O as a function of
$\xi_P/T$. All other ions and/or elements contribute less than 10\% to $\Delta
H$. 

On the left the solid and the dotted-and-dashed curves respectively correspond
to a WA ionized by an SED with $T_{in} = 10 \,\, \rm{and} \,\, 30 \ev$ (see
\eqn{GenSpectra}). In both cases $\eta = 0$ so that there is no {\it soft
excess} component in these SEDs, so any changes in the heating fractions are
entirely due to the variation in the accretion disk spectrum. 

On the right panels the solid and the dotted-and-dashed curves correspond to
$\eta = 0 \,\, \rm{and} \,\, 3$ respectively, while $T_{in} = 10 \ev$. Hence
changes in the heating fractions in the right panels are due only to the
varying strength of the {\it soft excess} component.

The top panels (labeled A) of \fig{Heating} show the region of the stability
curves where the variation of these spectral components has maximum effect.

\subsubsection{Hydrogen and Helium}
\labsubsubsecn{HHe}

Panels B and C of \fig{Heating} show the distribution of $\Delta H$
respectively for neutral Hydrogen H$^{+0}$ and singly ionised Helium He$^{+1}$.
In the left panels we see that there is no significant difference in the
distributions as $T_{in}$ increases from $10 \,\, \rm{to} \,\, 30 \ev$. For
both H$^{+0}$ and He$^{+1}$, $\Delta H$ is considerably less for $\eta = 3.0$
than for $\eta = 0$ as seen in the right panels. Thus in either case (of the
disk blackbody or the {\it soft excess}) H$^{+0}$ and He$^{+1}$ can be ruled
out as the cause of extra heating causing the increase in the temperature of
the stability curves in the range of $\log(\xi_P/T)$.

\subsubsection{Iron}
\labsubsubsecn{Iron}

We find that in the disk blackbody and the {\it soft excess} cases  the
significant heating agents are the various species of iron. It is to be noted
that the range of $\Delta H$ for Fe (panels D1, D2 \& D3) is double that for O
(panels E1, E2 \& E3). The $\Delta H$ behaviour of the iron ions are affected,
both, when we increase $T_{in}$ from $10 \,\, \rm{to} \,\, 30 \ev$, and when we
increase $\eta$ from 0 to 1.  $\rm{Fe^{+7}, Fe^{+8}, \,\, Fe^{+9}, \,\,
Fe^{+10}}$ and the Unresolved Transition Array (UTA, see below) are the most
significant heating agents; and their contributions have been added to give the
heating fraction due to iron (labeled `Iron') in panels D1 through D3. On the
left panels D2 and D3 the heating fractions of the individual ions of iron and
the UTA are shown, respectively, for $T_{in} = 10 \,\, \rm{and} \,\, 30 \ev$.
Similarly, the right panels D2 and D3 show $\Delta H$ for the individual iron
ions and the UTA, respectively, for $\eta = 0 \,\, \rm{and} \,\, 3.0$. A WA
which is overabundant in iron \citep{fields05, fields07} and/or illuminated by
a SED having photons facilitating absorption by iron will be warmer at
relatively lower values of $\xi_P$. These two physical scenarios influence
different ions of iron, which may help to distinguish between the otherwise
degenerate effects.
 
\begin{figure*}
\begin{center}
\includegraphics[scale = 1, width = 0.85\textwidth, angle = 0]{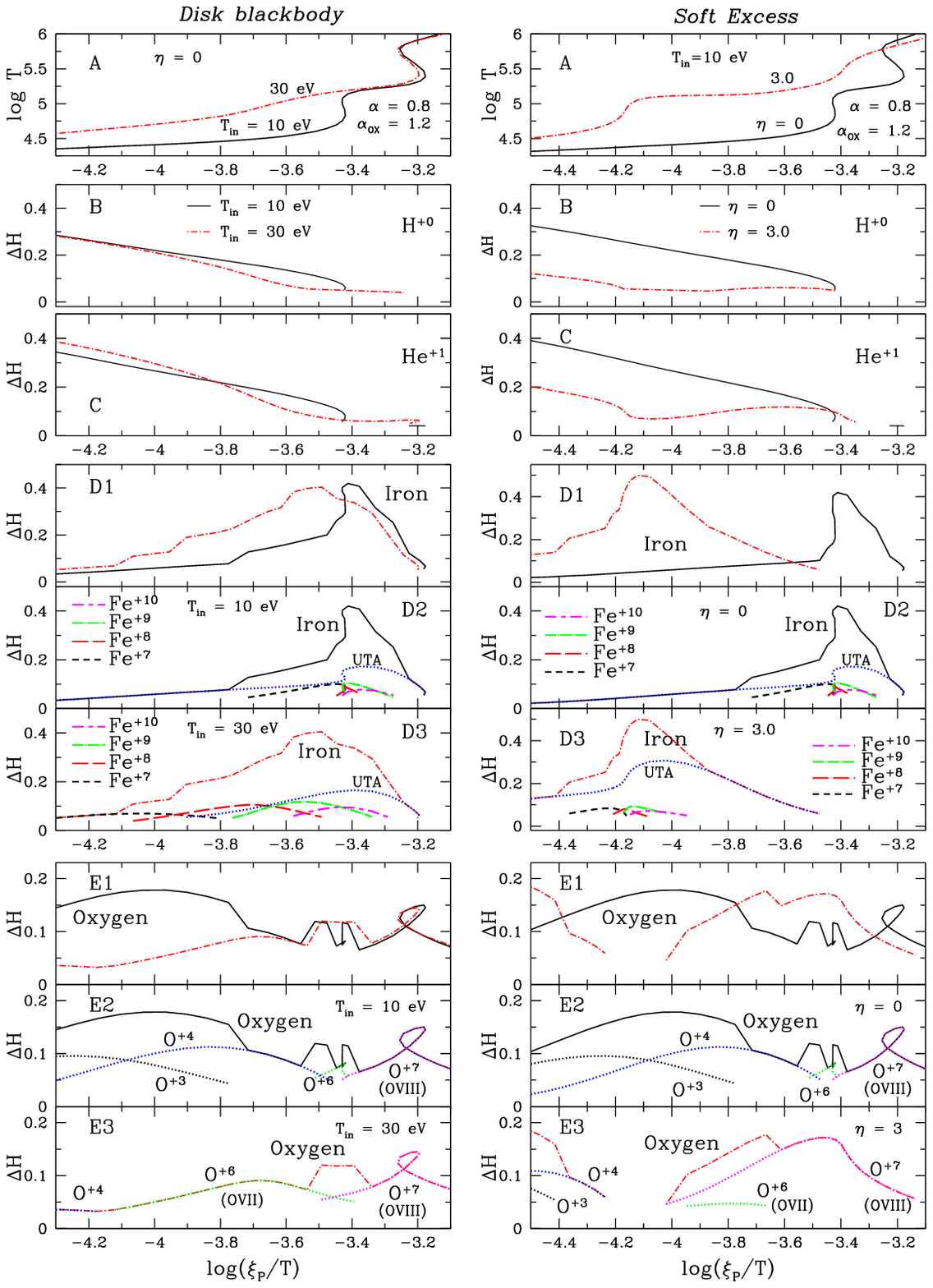}
\caption[The principal heating agents responsible for forming stable WA]
{The principal heating agents which are being influenced by the soft spectral
components {\it disk blackbody} (panels on the left) and {\it soft excess}
(panels on the right) present in the SEDs of typical AGN. \emph{Top panels (A)}
: The stable part of the thermal equilibrium curves for an optically thin Solar
metallicity gas illuminated by ionizing continua given by \equn{GenSpectra}. On
the left panel the solid and the dotted-and-dashed lines correspond to SEDs
having $T_{in} = 10 \,\, \rm{and} \,\, 30 \ev$ respectively with $\eta = 0$ in
both cases. The solid line in the right panel represents a SED with $\eta = 0$
and the dotted-and-dashed curve is for $\eta = 1.0$, where $T_{in} = 10 \ev$
for both the curves. We have zoomed in on the regions of the stability curves
where they differ most from one to the other. The line schemes described for
the left and right panels are maintained same for the lower panels in the
figure. \emph{Lower panels} : Fraction ($\Delta H$) of the total heating caused
by the various significant elements and ions. We have demonstrated the effects
of all those elements which contribute $\gtrsim 10\%$ to the total heating in
the same $\log(\xi_P/T)$ range as shown in the top panels. Panels B, C, D1 and
E1 show the results for H$^{+0}$, He$^{+1}$, iron and oxygen respectively. The
`break-up' in heating fraction of iron are shown in panels D2 and D3 and the
same for oxygen are shown in panels E2 and E3. Note that the y-range for the
panels E1 through E3 for Oxygen are different from all other panels showing
$\Delta H$ distributions because Oxygen never contributes more than 20\% to the
heating unlike other agents, some of which are responsible for even up to 50\%
of the heating for some values of $\log(\xi_P/T)$. Note that the discontinuity
in the $\Delta H$ distribution of oxygen in the $\eta = 3.0, T_{in} = 10 \ev$
case (dotted-and-dashed curve in panels E1 and E3) is because the contribution
drops below 5\% for $-4.2 < \log (\xi_P/T) < 4.0$.} 
\labfig{Heating}
\end{center} 
\end{figure*}


\begin{figure*}
\begin{center}
\includegraphics[scale = 1, height = 0.8\textwidth, trim = 20 100 20 100 , clip, angle = 90]{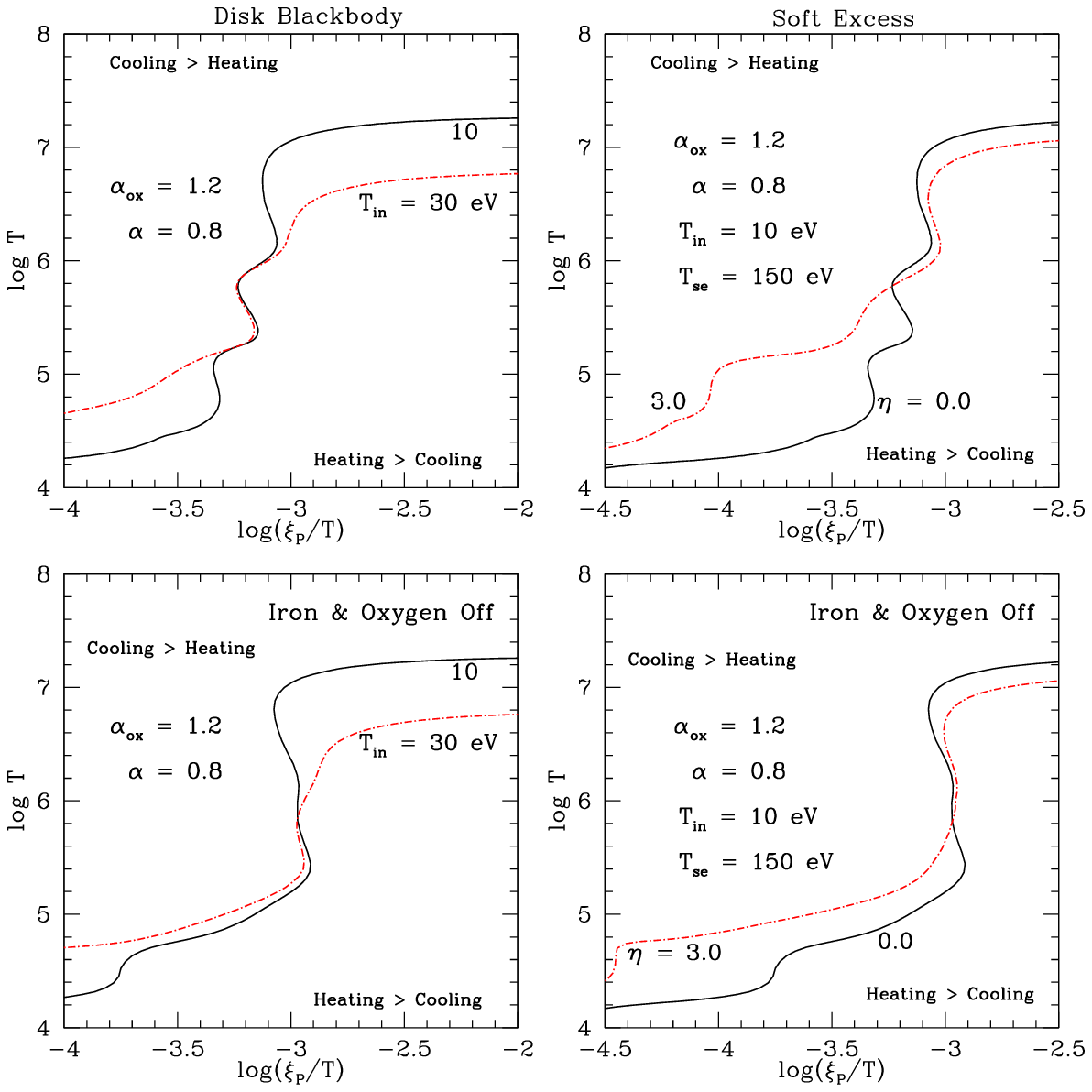}
\caption[]
{\emph{Top Panels :} The comparison between the stability curves for Solar
metallicity gas, when ionized by SEDs having accretion disks with different
temperatures $T_{in} = 10 \,\, \rm{and} \,\, 30 \ev$ (left) and when the
ionizing continua have different strengths of the {\it soft excess} component
$\eta = 0 \,\, \rm{and} \,\, 3.0$ (right). \emph{Bottom panels :} The same
ionizing continua are used to illuminate a gas for which iron and oxygen have
been removed. The relative differences in the stability curves are almost
entirely removed in the lower left panel, and are significantly reduced in the
lower right panel, indicating that these spectral components strongly influence
the ions of iron and oxygen.}  
\labfig{IronOxygOff}
\end{center}
\end{figure*}

For $-4.0 < \log(\xi_P/T) < -3.4$, where $\Delta H$ for iron is most affected
by $T_{in}$, Fe$^{+8}$ and Fe$^{+9}$ are the dominant ions (panel D2 and D3 on
the left of \fig{Heating}). The ionization potentials for $\rm{Fe^{+7},
Fe^{+8}, \,\, Fe^{+9}, \,\, \rm{and} \,\, Fe^{+10}}$ are 151.1, 235, 262.1 and
290.4 $\ev$ respectively. As the accretion disk becomes hotter, with $T_{in}$
increasing from 10 to 30 eV, the number of photons with $E > 100 \ev$ increases
by a factor of 27. It is this enhancement in the number of photons that
influences these ions of iron differently in the two cases and hence changes
the shape of the stability curves.

The UTA is a blend of the numerous absorption lines arising from the iron
M-shell ions, Fe$^{+0}$ - Fe$^{+15}$, due to their n = 2-3 (mainly 2p-3d, n
being the principal quantum number of the active electron) transitions which
are found to occur between 16 and 17 $\AA$ (730 - 776 $\ev$) \citep{behar01,
sako01, netzer04, krongold05a, krongold05b, holczer07}.  The energy ranges
important for the UTA are likely to be influenced by the {\it soft excess}
component of the SED. This effect is clearly seen by comparing the $\Delta H$
distributions of UTA in the right panels D2 and D3 of \fig{Heating}; $\Delta H$
for iron is far more dominated by the UTA $\Delta H$ contribution for $\eta =
3.0$ case in the range $-4.4 < \log(\xi_P/T) < 3.6$.

Note that in panels D2 (both left and right), the solid curve for iron is
multivalued at $-3.43 < \log(\xi_P/T) < -3.42$, just where the stability curve
is multivalued (see panels A), giving multiple phases in pressure equilibrium.
On the other hand, for the $T_{in} = 30 \ev \,\, \rm{and \,\, the} \eta = 3$
curves (the dotted-and-dashed stability curves in panels A and the
corresponding $\Delta H$ distribution of iron in panels D3, both left and
right) there is no multivalued behaviour for $\log(\xi_P/T) \sim -3.4$. Thus
the behaviour of the stability curve seems to be driven by the heating due to
the different ions of iron.

\subsubsection{Oxygen}
\labsubsubsecn{Oxygen}

Panels E1 through E3 show the results for the heating fractions contributed by
Oxygen. Note the factor 2 smaller, y-range for the panels E1 through E3 because
Oxygen never contributes more than 20\% to the heating of the gas.

The left panels show that $\Delta H$ due to Oxygen is lower for $T_{in} = 30
\ev$ than that for $T_{in} = 10 \ev$ in the range $\log (\xi/T) \lesssim -3.6$.
In the same range 0f $\log (\xi/T)$, the stability curve for $T_{in} = 30 \ev$
has higher temperature than the $T_{in} = 10 \ev$ curve (panel A on the left).
Thus here Oxygen is not acting as one of the required heating agents resulting
in a warmer absorber. 

However, in the $\eta = 3.0$ case (right panels) Oxygen becomes a significant
excess heating agent for $\log(\xi_P/T) > -3.6$ where the contributing ions are
O$^{+6}$ (OVII) and O$^{+7}$ (OVIII).

As for iron, the $\Delta H$ distribution due to oxygen for the $T_{in} = 10
\ev, \eta = 0$ case, are also multivalued at the same values, namely the narrow
range $-3.43 < \log(\xi_P/T) < 3.42$ (panels E2, both left and right), although
not as strongly as that for iron (panels D2, both left and right). Moreover, at
$\log(\xi_P/T) \sim -3.4$, it is iron and not oxygen which is the dominant
heating agent. However, oxygen is the dominant heating agent at higher values
$\log(\xi_P/T) \sim -3.2$, and has multivalued $\Delta H$ distribution which
drives the multi-phase nature of the solid stability curves at those values of
$\log(\xi_P/T)$. 

The $\Delta H$ distribution of oxygen for the $T_{in} = 30 \ev, \eta = 0$ is
shown in the left E3 panel (dotted-and-dashed curve). For $-3.8 < \log(\xi_P/T)
\lesssim -3.3$ the oxygen $\Delta H$ is about 2 to 4 times lower than the iron
$\Delta H$ (left panel D3). The oxygen $\Delta H$ dominates only beyond
$\log(\xi_P/T) > -3.3$ where it is multivalued and is responsible for the
multivalued nature of the stability curve (left panel A) at the same values of
$\log(\xi_P/T)$.    

For the $\eta = 3.0, T_{in} = 30 \ev$ case (right panel E3) as well, the oxygen
$\Delta H$ contribution is lower than the iron $\Delta H$ (right panel D3)
until $\log(\xi_P/T) \gtrsim -3.6$ beyond which the oxygen $\Delta H$
distribution determines the behaviour of the stability curve. Note that the
stability curve (dotted-and-dashed in right panel A) is not multivalued at
$\log(\xi_P/T) \gtrsim -3.6$ because the $\Delta H$ for oxygen is not
multivalued.

Thus, at the higher ionization ($\log(\xi_P/T) \sim -3.2$) phase of the WA, the
nature of the stability curve seems to be driven by the heating due to the
oxygen ions, predominantly O$^{+7}$ (OVIII).

\subsubsection{Zero oxygen and iron abundance}
\labsubsubsecn{ElementsOff}

\begin{figure*}
\begin{center}
\includegraphics[scale = 1, height = \textwidth, trim = 110 100 125 100, clip, angle = 90]{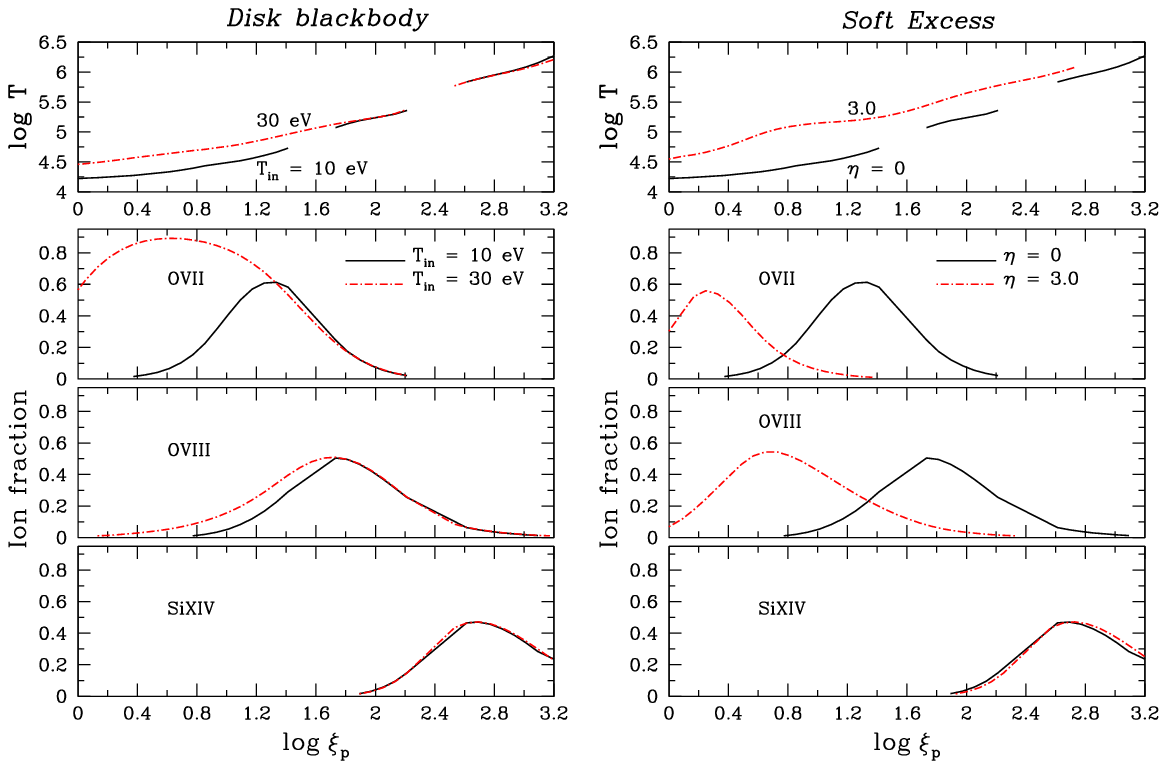}
\caption[The ion fractions of some of the relevant ions as a function of ionization parameter.]
{The ion fractions of some of the relevant ions as a function of ionization
parameter; the left panels show how this distribution is influenced by the {\it
disk blackbody} spectral component and the right panels show the same for the
{\it soft excess}. \emph{Top panels} : The distribution of temperature as a
function of $\xi_P$ is plotted for the thermally stable WA. The gaps in the
curves correspond to the range of $\xi_P$ over which the stability curve is
thermally unstable. An increase in accretion disk temperature $T_{in}$ from 10
to 30 $\ev$ makes the gas stable at $\log T \sim 5$. When the strength of the
{\it soft excess} is increased from $\eta = 0 \,\, \rm{to} \,\, 3.0$, all the
phases of the WA become thermodynamically stable up to a temperature of $\log T
\sim 6.3$  \emph{Lower panels} : Ion fraction for various ions (as labeled)
important for the WA states. The line-style scheme followed here is the same as
in \fig{Heating} and is also labeled in the second panel from the top. See text
for the physical implications of the ion fraction distribution.}  
\labfig{IonFrac}
\end{center}
\end{figure*}

We have further illustrated the importance of iron and oxygen as the key
players in determining the thermal properties of the WA using
\fig{IronOxygOff}. The distinction between the left and the right panels are as
that in \fig{Heating}; the upper panels being the same stability curves as in
\fig{Heating} panels A, but for slightly different ranges of $\log(\xi_P/T)$
and wider ranges of $\log T$.

Comparing the stability curves for Solar metallicity gas (top panels), which
have pronounced kinks in the temperature range $4.2 < \log T < 6.5$, with the
relatively featureless ones for gas with zero O and Fe (bottom panels), we see
that the  detailed shape of the curves in the temperature range $\sim 4.2 <
\log T < 6.5$ are governed by the atomic interactions due to oxygen and iron
\citep[also see][]{chakravorty09}. As a result, the SED induced differences
between the solid and the dotted-and-dashed stability curves in the top panels
are significantly smoothed out in the bottom panels.

For $-3.7 < \log (\xi_P/T) < -3.3$ the rise in temperature of the Solar
metallicity stability curves in the top left panel shows a temperature
difference of 0.4 dex from one curve to the other. This difference is reduced
to 0.2 dex when iron and oxygen are not present in the gas (lower left panel)
suggesting that a hotter accretion disk ($T_{in} = 30 \ev$) mainly affects ions
of iron and oxygen which results in a hotter WA. 

The temperatures of the Solar metallicity stability curves in the top right
panels are remarkably different, by $\sim 1$ dex, for $-4.2 < \log (\xi_P/T) <
-3.3$. These differences are significantly reduced to 0.3 - 0.5 dex when iron
and oxygen are removed from the gas (lower right panel).  This indicates that
atomic interactions due to these elements are affected by the strength of the
{\it soft excess} component. Unlike the accretion disk case, however, the
stability curves in the lower right panel retain some of their differences in
WA temperatures suggesting that other elements have a role as heating agents in
the {\it soft excess} case. The detailed investigation of this effect is beyond
the scope of this paper, but will be attempted soon in our future publications.  



\subsection{Ion fraction}
\labsubsecn{HeatingCooling_IonFrac}

Absorption lines and edges of OVII, with ionisation potential (IP) of 0.74 keV,
and OVIII, with IP of 0.87 keV, are often prominent signatures of the WA in
soft X-ray spectra. Hence the column densities of these two ions are often
considered as important observable parameters for WA states (see introduction
for references). At higher values of $\xi_P$, absorption by various ions of
silicon become important, e.g. \citet{netzer03} show that SiXIV (IP of 2.67
keV) has a significant column density corresponding to the high temperature
component of the WA. We choose these three ions to study the variation of their
ion fraction as a function of $\xi_P$ as the spectral components change in the
ionizing continuum.

The ion fraction $I(X^{+i})$ of the $X^{+i}$ ion is the fraction of the total
number of atoms of the element $X$ which are in the $i^{\rm{th}}$ state of
ionization. Thus, $$I(X^{+i}) = \frac{N(X^{+i})}{f(X) \, N_{\rm{H}}},$$ where
$N(X^{+i})$ is the column density of the $X^{+i}$ ion and $f(X) =
n(X)/n_{\rm{H}}$ is the ratio of the number density of the element $X$ to that
of hydrogen.

The top panels of \fig{IonFrac} show the distribution of $\log T$ with respect
to $\log \xi_P$ for the stable WA phases. The gaps in the lines correspond to
the range of ionization parameter over which the gas is thermally unstable. The
lower panels show I(OVII), I(OVIII) and I(SiXIV).

As before, left panels in \fig{IonFrac} correspond to the changes in the
diskblackbody. For $T_{in} = 10 \ev$ we can see three distinct regions of
$\xi_P$ separated by intermediate ranges of thermally unstable solutions. These
regions are respectively dominated by OVII, OVIII and SiXIV.  However, when
$T_{in}$ is raised to $30 \ev$, the stability curve becomes stable all the way
up to $\log T = 5.3$ and the OVII dominated phase gradually merges in to the
OVIII dominated gas, as demonstrated by the uninterrupted dotted-and dashed
line in the top left panel. Thus there seems to be a continuous distribution of
possible thermal solutions resulting from the greater production of OVII
(second panel from top) which, in its turn, is a result of the enhanced number
of $E \gtrsim 100 \ev$ photons coming from a hotter accretion disk with $T_{in}
= 30 \ev$. The change in the shape of the {\it disk blackbody} does not change
the ion fraction distribution of the high ionization species like OVIII or
SiXIV (the two lowest panels) or the stability curve for $\log T > 5$ and $\log
\xi_P > 1.7$. Thus the higher temperature ($\log T \gtrsim 5$) phase of the
ionized absorber remains unaffected.  

The effect of the {\it soft excess} component is demonstrated in the right panels of \fig{IonFrac}. 
In this case the increase in the strength of the {\it soft excess} component 
from $\eta = 0$ to $\eta = 3.0$ makes the absorber thermally stable all the way up to temperatures 
of $\log T \sim 6$. The third panel from the top shows that increase in the strength of the 
{\it soft excess} component mainly influence the occurrence of OVIII. The fraction of OVII is reduced for $\eta = 3.0$ because of 
the facilitated production of OVIII at lower values of $\log\xi_P$. This 
is caused by increase in the number of $\sim 100 \ev - 2 \kev$ photons by a factor of $\sim 7.5$. 
The ion fraction distribution of SiXIV however, remains unaffected (bottom panel).


\section{Other significant physical parameters from earlier studies}
\labsecn{OtherPars}

\citet{chakravorty09} had studied the properties of the WA as a function of the X-ray spectral 
index $\alpha$ and the chemical abundance of the gas. In this subsection we summarise some of those 
results which are relevant in the context of the soft spectral components of typical AGN continua and 
their effect on the WA. 

\subsection{X-ray spectral index $\alpha$}
\labsubsecn{Alpha}

\citet[][Section 5.1 and Figure 3]{chakravorty09} showed that there was no
thermally stable absorber for $\log T > 4.4$ for a flat spectral index of
$\alpha = 0.2$ (in \equn{2powerlaw}, $\alpha_{OX} = 1.2$). \fig{Alpha} shows
the stability curve using \equn{2powerlaw} with $\alpha = 0.2$ (dotted, magenta
line), which corresponds to an ionizing continuum with no {\it soft excess}
component, shows no appreciable stable WA phase at $T \sim 10^5 \kel$. In
\secn{Scurve_Se} of this paper we have seen that the presence of {\it soft
excess} component in the AGN spectra facilitates the existence of thermally
stable gas at $\log T \sim 5.0$. However, throughout we have used $\alpha =
0.8$. It is thus crucial to check if the {\it soft excess} component can result
in stable $10^5 \kel$ gas even if $\alpha \sim 0.2$. 

\begin{figure}
\begin{center}
\includegraphics[scale = 1, width = 8 cm]{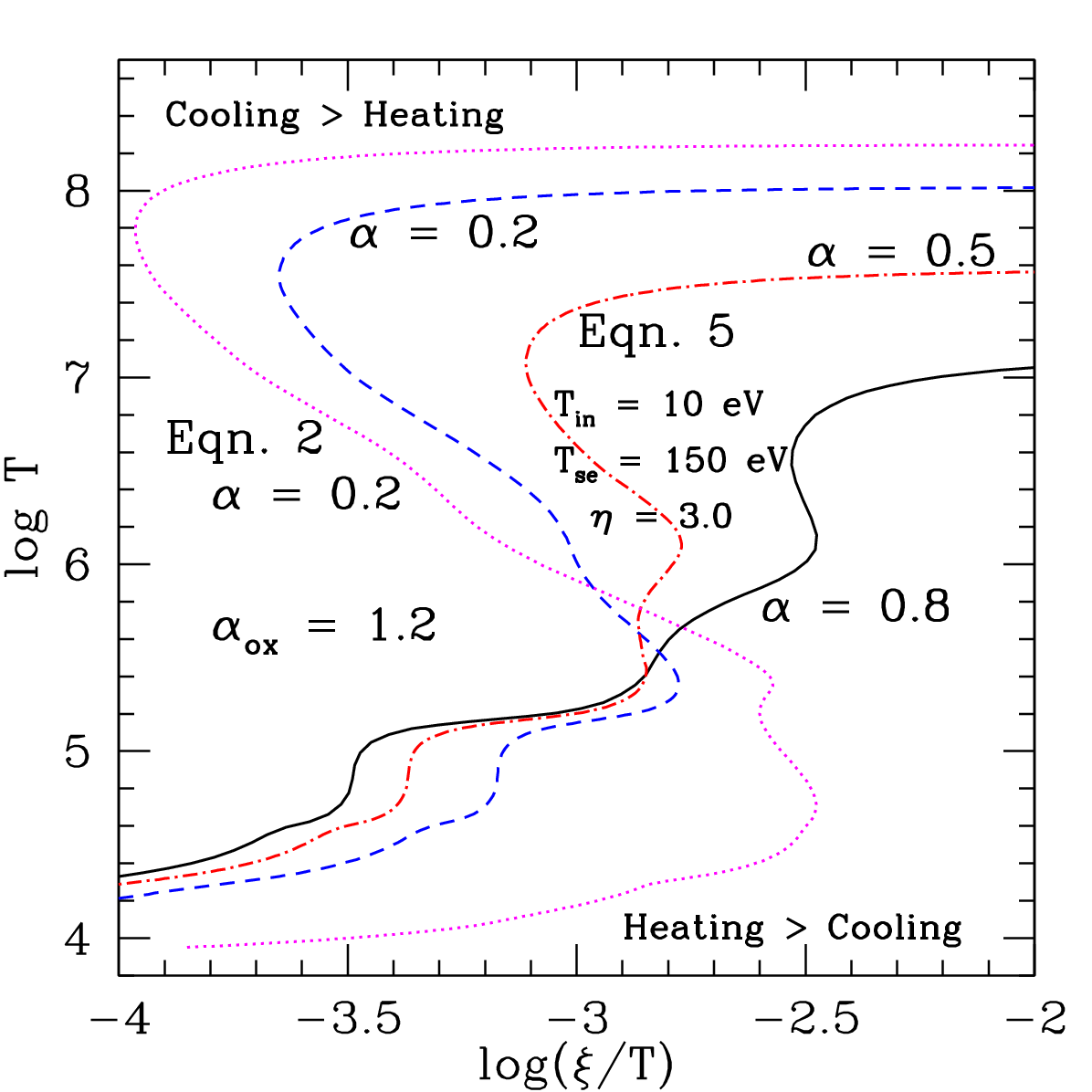}
\caption[]
{Thermal equilibrium curves generated using SEDs given by \equn{GenSpectra}
with a strong {\it soft excess} ($T_{se} = 150 \ev$, $\eta = 3.0$ and $T_{in} =
10 \ev$) for a range of X-ray slope $\alpha = 0.8 \,\, 0.5 \,\, \rm{and} \,\,
0.2$ (respectively drawn with the solid, the dotted-and-dashed and the dashed
lines). The stable phase at $\log T \sim 5.0$ the range of $\log(\xi/T)$
progressively decreases as $\alpha$ decreases from 0.8, to 0.5 and then to 0.2.
For comparison we have used \equn{2powerlaw} with $\alpha = 0.2$, which
corresponds to an ionizing continuum which has no {\it soft excess} component,
and drawn the dotted stability curve which is consistent with no stable WA
phase at $\log T \sim 5.0$.}  
\labfig{Alpha}
\end{center}
\end{figure}

The solid, the dotted-and-dashed and the dashed curves in \fig{Alpha} show the
comparison between the stability properties of the WA ionized by SEDs given by
\equn{GenSpectra} with $T_{se} = 150 \ev$, $\eta = 3.0$ and $T_{in} = 10 \ev$
but various values of $\alpha$ (0.8, 0.5 and 0.2). The range, in $\log(\xi/T)$,
of stable WA phase at $\log T \sim 5$ decreases as a function of decreasing
value of $\alpha$. However, even for $\alpha$ as low as 0.2 we still see a
stable phase at $\log T \sim 5$. Thus the appearance of a stable state at $\log
T \sim 5.0$ for the $\alpha = 0.2$ curve can be attributed to the presence of a
strong {\it soft excess} component present in the ionizing continuum. 
However, if $\alpha \lesssim 0.2$, even with contributions from a strong {\it
soft excess} component, there is no evidence for stable gas at any higher
temperature than $\log T > 5.3$ (refer to the dashed, blue curve). 

Notice that in \fig{Alpha} the stability curves are drawn using $\xi$ instead of 
$\xi_P$. The normalisation scheme for the ionisation parameter discussed in \subsecn{XiP} 
is not appropriate while comparing SEDs with different values of the X-ray spectral index 
$\alpha$.


\subsection{Chemical abundance of the absorber}

We discuss the role of the elemental abundances of the absorbing medium in
\citet{chakravorty09}.  Here, we revisit the super-Solar metallicity results,
but, with more realistic AGN continua given by \equn{GenSpectra} ($\alpha =
0.8$, $\eta = 3.0$ and $T_{in} = 10 \ev$). The stability curves are shown in
\fig{Metals} for WA abundances from Solar (Z$_{\odot}$) to 9 Z$_{\odot}$ in
steps of 2. The qualitative results are the same as in \citet{chakravorty09}. 

On the low temperature ($\log T \lesssim 4.5$, corresponding to OVII like ions)
arm of the stability curves, super-Solar metallicity results in a cooler
absorber for the same $\xi_P$ (or $\xi$) values, with an increase in the range
of $\xi_P$ over which we get stable WA. Thus super-Solar metallicity opposes
the role of increased $T_{in}$ (see \fig{DbbSpectraScurves} in
\subsecn{Scurve_Dbb}). 

For the intermediate temperature  arm of the stability curve ($\log T \sim
5.0$, corresponding to OVIII like ions) super-Solar abundance decreases the
range of stable WA, thus opposing the influence of the increase in $\eta$ in
the ionizing continuum (see \fig{BbSeSpectraScurves} in \secn{Scurve_Se}). 

On the highest temperature arm of the WA ($\log T \sim 6.0$, corresponding to
SiXIV like ions) super-Solar gas tends to be hotter, and is thermodynamically
stable for a larger range of values of $\log(\xi_P/T)$. Thus the fine tuning of
$\xi_P$ values to detect such species of ions is relaxed if the absorber has
super-Solar abundance, and the probability of detecting these high ionization
species is increased. In this regime, the influence of high metallicity is
degenerate with the influence of increasing $\alpha$ in the ionizing continuum,
but only if $\alpha \gtrsim 0.8$ (see \fig{Alpha} in \subsecn{Alpha}).

\begin{figure}
\begin{center}
\includegraphics[scale = 1, width = 8 cm]{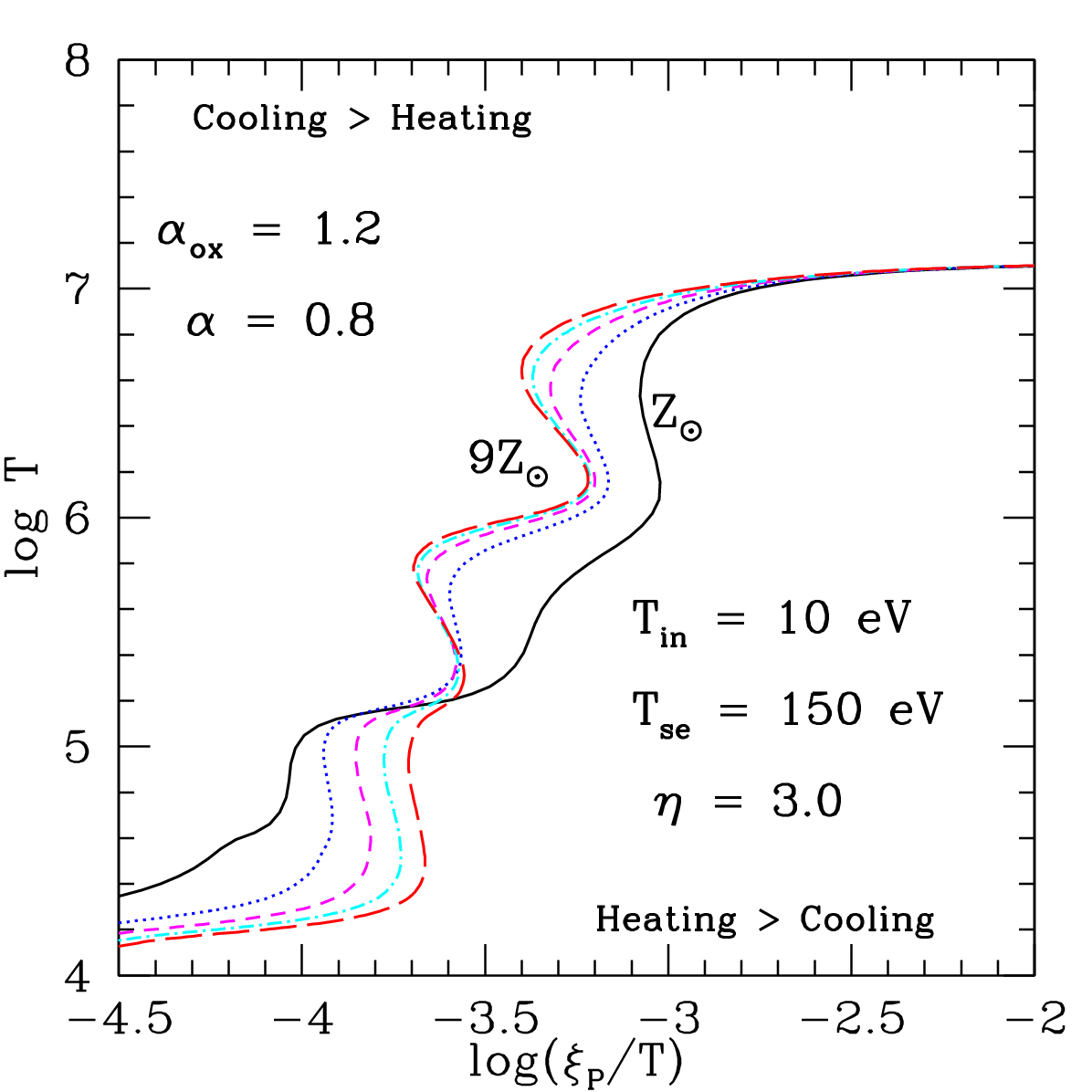}
\caption[Stability curves for absorbers with super-Solar abundance.]
{Stability curves for absorbers with Solar and super-Solar abundance from
Z$_{\odot}$ to 9 Z$_{\odot}$ in steps of 2. The ionizing continuum has a strong
soft excess (\equn{GenSpectra} with : $\alpha = 0.8$, $\eta = 3.0$ and $T_{in}
= 10 \ev$).}  
\labfig{Metals}
\end{center}
\end{figure}


\section{Multiphase warm absorber}
\labsecn{Multi-phase}

The stability curves we have derived in the previous sections often contain
segments that allow phases at different temperatures to occur at similar
pressure $\xi_P/T$ indicating the possibility of two or more distinct WA phases
within the same medium to exist at near pressure equilibrium. However,
variations in some of the physical parameters can also result in the
possibility of multi-phase absorbers being lost. 

Observationally, distinct phases in pressure equilibrium have been derived in a
number of cases of WAs \citep{krongold03, netzer03, chelouche05, krongold07,
andrade-velazquez10}, while in other cases it has been claimed that there is a
continuous range of ionization parameter \citep{ogle04, steenbrugge05}. Since a
wide range of physical parameters have been studied in this paper and
\citet{chakravorty09} for their effects on the WA, we can now form a clearer
picture of which physical conditions favour the multi-phase absorbers.

The {\it disk blackbody} component from the accretion disk affects the lower
temperature part of the stability curve (\dsecn{Scurves}{HeatingCooling}).
\fig{DbbSpectraScurves} shows that the stability curve (solid line) with
$T_{in} = 10 \ev$ has a slim range $-3.34 \le \log(\xi_P/T) \le -3.31$ over
which gas at low temperatures $\log T \lesssim 4.3$ can be in pressure
equilibrium with the absorber at intermediate temperatures $\log T \lesssim
5.0$. However, as $T_{in}$ is increased, all possibilities of such multi-phase
structure is lost, instead the curve becomes stable in the intermediate
temperature range supporting a continuous distribution of allowed pressure and
temperature in this phase space. Thus cooler accretion disks support a
multi-phase WA, although for a narrow range of pressure.   

An ionizing continuum having no {\it soft excess} (solid line in
\dfig{BbSeSpectraScurves}{CompSeSpectraScurves}) allows for a multi-phase WA 
between $\log T \sim 5 \,\,\rm{and} \,\,6$. However, with the increase in the
strength of the {\it soft excess} component, the range of $\log(\xi_P/T)$, over
which the possibility of multi-phase WA exists reduces from 0.09 dex for $\eta =
0$ to 0.006 dex for $\eta = 1.0$ and to 0 for $\eta = 3.0$.

\citet{chakravorty09} had emphasised that super-Solar abundances increase the
range of $\log(\xi/T)$ with the possibility of multi-phase WA \citep[also
see][]{komossa01}. For example, as the metallicity of the gas was increased
from 1.0 to 3.0 to 5.0, $\Delta_{45}[\log(\xi/T)]$ increased from 0.012 to 0.08
to 0.1, where $\Delta_{45}[\log(\xi/T)]$ gives the range of common values of
$\log(\xi/T)$ over which the WA has stable phases at both $\sim 10^5$ and $\sim
10^4 \kel$.  Beyond a metallicity of 5Z$_{\odot}$, $\Delta_{45}[\log(\xi/T)]$
was found to drop because the range of $\log(\xi/T)$ for the $\sim 10^5$K WA was
found to decrease with the increase in metallicity and became zero for a gas
with metallicity 9Z$_{\odot}$. 

Similar trends are retained for the continua in this paper. As the metallicity
of the gas is increased, keeping the ionizing continuum the same, the stable
phases increasingly align along the same values of $\log(\xi_P/T)$
(\fig{Metals}), strongly suggesting the presence of multi-phase WA.
$\Delta_{45}[\log(\xi_P/T)]$ increases from 0 to 0.02 to 0.04 to 0.045 as
metallicity is increased from Solar to 3, 5 and 7 times Solar.  Beyond this
high metallicity, $\Delta_{45}[\log(\xi_P/T)]$ becomes constant.  There is a
even more significant increase in $\Delta_{56}[\log(\xi_P/T)]$ from 0 to 0.03
to 0.08 to 0.11 to 0.13 with the increase in metallicity from Solar to 3, 5,7
and 9 times Solar.	

For NGC 3783, \citet{netzer03} found three distinct phases of the WA in near
pressure equilibrium. In \fig{Metals} we see that only for Z = 9Z$_{\odot}$
there is a very narrow range of 0.03 dex in which all three phases of the WA,
at temperatures $\sim 10^4, 10^5 \,\, \rm{and} \,\, 10^6 \kel$, are stable in
the range $-3.69 \le \log(\xi_P/T) \le -3.66$. Interestingly, the stability
curve for a much lower metallicity, Z = 5Z$_{\odot}$, gas but ionized by a
continuum with weaker {\it soft excess} ($\eta = 1.0$, not shown in
\fig{Metals}) shows the three phases in pressure equilibrium for a larger range
of 0.08 dex for $-3.48 \le \log(\xi_P/T) \le -3.4$. Thus, it is easier
(relatively lower super-Solar metallicity) to have 3 phases in pressure
equilibrium if the strength of the {\it soft excess} is lower.   In all the
physical conditions of the WA, that were studied by \citet{chakravorty09},
three phases of the WA were found to coexist in pressure equilibrium for still
lower metallicity $\gtrsim 3\zsol$ which brings it down to observed values
\citep[e.g.  see][for MRK 279]{fields05,fields07}. It is to be noted, however,
that the continua used by \citet{chakravorty09} had no {\it soft excess}
component in them.


\section{Conclusion}
\labsecn{Conclusion}

We have examined the effect of spectral energy distributions including hot
disks and {\it soft excess} spectral components in the energy `blind spot'
between 13 - 100 eV (\dequn{GenSpectra}{CompSe}) on warm absorber. We
investigated whether the spectral shapes can, in turn, be constrained from the
observed properties of the warm absorber.

We summarise our results as follows :
\begin{enumerate}
\item[$\bullet$] The maximum temperature of the accretion disk component (see
\equn{GenSpectra}, \subsecn{SED_Se}) strongly affects on the low temperature
($\log T \lesssim 4.5$) arm of the stability curve, the thermal properties of
which are largely decided by ions Fe$^{+7}$ to Fe$^{+10}$ and O$^{+6}$ (OVII).
This phase of the WA becomes hotter and thermodynamically stable over a larger
range of $\xi$ when the ionising continuum is hotter ($T_{in} = 30 \ev$)
compared to $T_{in} = 10 \ev$. Thus hotter accretion disks are more likely to
produce WA phases characterised by ions having similar ionisation potentials to
OVII. However, the possibility of this phase of the WA being in pressure
equilibrium with the higher temperature phases is eliminated in the hot disk
cases as $\Delta_{45}[\log(\xi/T)]$ decreases from 0.03 for $T_{in} = 10 \ev$
to 0 for $T_{in} = 20 \,\, \rm{and} \,\, 30 \ev$. Changes in the accretion disk
temperature however, do not affect the higher temperature arms of the stability
curve.
\item[$\bullet$] The thermal properties of the intermediate temperature ($\log
T \sim 5.0$) branch of the stability curve are essentially determined by atomic
interactions due to the different ions of iron and OVIII (IP = 0.87 keV). We
find that with the increase in the relative strength $\eta$ of the {\it soft
excess} component this $10^5 \kel$ stable phase of the WA spans a much larger
range of $\log(\xi/T)$ (by 0.4 dex from $\eta = 0$ to $\eta = 3.0$) thus
increasing the probability of finding WA phases characterised by `OVIII like'
ions.  
\item[$\bullet$] For a gas whose chemical composition is devoid of iron and
oxygen, changes in the accretion disk spectral component bring about no change
in the properties of the WA, and changes due to the variation in the strength
of the {\it soft excess} component are significantly reduced.
\item[$\bullet$] The highest temperature arm ($\log T \sim 6.0$) of the
stability curve which is characterised by the ions having similar IP to that of
SiXIV (IP = 2.67 keV), is left almost unaffected by any changes either in the
accretion disk component or in the {\it soft excess} component.
\item[$\bullet$] An AGN continuum with a flat X-ray slope will not produce high
IP ions like SiXIV, and may show signatures of lower IP ions like OVIII (IP =
0.87 keV) only if a sufficiently strong {\it soft excess} component is present
in the SED.
\item[$\bullet$] The metallicity of the gas plays an important role in
determining the multi-phase nature of the warm absorber. The possibility of the
$10^4 \,\, \rm{and \,\, the} \,\, 10^5 \kel$ phases of the warm absorber
occurring at similar values of $\log(\xi/T)$ is increased if the absorber is
super-Solar in abundance. Similarly, the chances of having pressure equilibrium
between the $10^5 \,\, \rm{and \,\, the} \,\, 10^6 \kel$ phases are also
increased by the super-Solar metallicity of the gas. However, all three phases,
together, are found to be in pressure equilibrium only if Z $> 3\zsol$, and the
required abundance to achieve this is increased if the ionizing continuum has
stronger {\it soft excess}. 
\end{enumerate}

\section*{acknowledgements}
SC sincerely thanks Prof. Andy Fabian and Prof. Hagai Netzer for providing
valuable suggestions in the early stage of preparing this paper.

\end{document}